\documentclass[aps,prx,twocolumn,superscriptaddress]{revtex4-2}
\usepackage[utf8]{inputenc}
\usepackage{graphicx}
\usepackage{comment}
\usepackage[english]{babel}
\usepackage{float}
\usepackage{capt-of}
\usepackage{booktabs}

\usepackage{amsmath}
\usepackage{amsfonts}
\usepackage{amssymb}
\usepackage{bm}
\usepackage{siunitx}

\usepackage{tikz}
\usepackage{xcolor}

\usepackage{mymacros}
\renewcommand{\selectlanguage}[1]{}

\usepackage{hyperref}
\hypersetup{
    colorlinks,
    citecolor=black,
    filecolor=black,
    linkcolor=black,
    urlcolor=black
}

\def\showefe{} 
\ifdefined\showefe
\newcommand{\efe}[1]{\textcolor{purple}{#1}}
\else
\newcommand{\efe}[1]{}
\fi

\def\showch{} 
\ifdefined\showch
\newcommand{\comch}[1]{\textcolor{red}{#1}}
\else
\newcommand{\comch}[1]{}
\fi

\newcommand{\myspace}{\,}

\begin{document}
\title{A Model for Self-Organized Growth, Branching, and Allometric Scaling of the Planarian Gut}
\author{Christian Hanauer}
\affiliation{Max Planck Institute for the Physics of Complex Systems, Dresden, Germany}

\author{Amrutha Palavalli}
\affiliation{Department of Tissue Dynamics and Regeneration, Max Planck Institute for Multidisciplinary Sciences, Göttingen, Germany}

\author{Baiqun An}
\affiliation{Department of Tissue Dynamics and Regeneration, Max Planck Institute for Multidisciplinary Sciences, Göttingen, Germany}

\author{Efe Ilker}
\email{ilker@pks.mpg.de}
\affiliation{Max Planck Institute for the Physics of Complex Systems, Dresden, Germany}

\author{Jochen C. Rink}
\email{jochen.rink@mpinat.mpg.de}
\affiliation{Department of Tissue Dynamics and Regeneration, Max Planck Institute for Multidisciplinary Sciences, Göttingen, Germany}
\affiliation{Faculty of Biology and Psychology, University of Göttingen, Göttingen, Germany}

\author{Frank Jülicher}
\email{julicher@pks.mpg.de}
\affiliation{Max Planck Institute for the Physics of Complex Systems, Dresden, Germany}
\affiliation{Center for Systems Biology, Dresden, Germany}
\affiliation{Cluster of Excellence Physics of Life, Technische Universität Dresden, Dresden, Germany}

\date{\today}
\begin{abstract}
The growth and scaling of organs is a fundamental aspect of animal development. However, how organs grow to the right size and shape required by physiological demands, remains largely unknown.
Here, we provide a framework combining theory and experiment to study the scaling of branched organs. As a biological model, we focus on the branching morphogenesis of the planarian gut, which is a highly branched organ responsible for the delivery of nutrients. Planarians undergo massive body size changes requiring gut morphology to adapt to these size variations. Our experimental analysis shows that various gut properties scale with organism size according to power laws.  
We introduce a theoretical framework to understand the growth and scaling of branched organs. Our theory considers the dynamics of the interface between organ and surrounding tissue to be controlled by a morphogen and illustrates how a shape instability of this interface can give rise to the self-organized formation and growth of complex branched patterns. Considering the reaction-diffusion dynamics in a growing domain representative of organismal growth, we show that a wide range of scaling behaviors of the branching pattern emerges from the interplay between interface dynamics and organism growth. Our model can recapitulate the scaling laws of planarian gut morphology that we quantified and also opens new directions for understanding allometric scaling laws in various other branching systems in organisms.
\end{abstract}

\maketitle
\newpage
\section{Introduction}
Hierarchically branching tubular structures are a common design principle of organs of multicellular organisms  \cite{wolpert_2019,gilbert_2010,raven_2013}. Examples include the tree-like ramifications of the airways of the vertebrate lung that provide a large surface area for efficient gas exchange, the vasculature of animals and plants that ensures high perfusion efficiency via hierarchical branching into vessels of progressively smaller diameters, or the network of ducts within the mammary gland of mammals that secrete milk for nourishment of the offspring \cite{ochoa-espinosa_2012,goodwin_2020}. Invertebrate examples of branched organs include the gut of planarian flatworms, which consists of three primary branches (hence their systematic designation as Tricladida; tri = three, cladida = branches). These primary branches sprout numerous secondary side branches that, via further branching, permeate nearly the entire body of the worms and give rise to the branched morphology of the organ \cite{ivankovic_2019, forsthoefel_2011}. The planarian gut simultaneously mediates food absorption and nutrient distribution and is therefore often referred to as gastro-vasculature. Invariably, branched organs form during development via the hierarchical and self-organized branching of precursor ducts and these processes are collectively referred to as branching morphogenesis \cite{davies_2006}. 
 
The self-organized formation of branched morphologies is also a common and well-studied phenomenon in non-equilibrium physics \cite{meakin_1998,godreche_1991,vicsek_1992,langer_1980}. Examples here include the formation of branched structures during solidification (e.g., snowflakes), viscous fluid flow, or dielectric breakdown.
Despite these seemingly different phenomena, a common principle underlies the formation of branched structures.
On an abstract level, all these phenomena can be described by the motion of an interface that separates two regions with different properties (e.g., solid and liquid regions in solidification). The formation of an interface protrusion increases the protrusion's growth rate, and this shape instability can lead to the formation of highly branched structures \cite{saffman_1958,mullins_1963}. Although shape instabilities can explain the formation of branched structures in a wide class of pattern forming systems, including bacterial colonies \cite{ben_jacob_2000}, it remains unclear if or how shape instabilities may contribute to the morphogenesis of branched animal organs.

A central regulatory principle in biological branching morphogenesis are so-called morphogens \cite{lu_2008,nelson_2009}, which are signaling molecules that spread within the tissue environment and influence cell behavior in a concentration-dependent manner \cite{gilbert_2010}. Morphogens are typically produced in dedicated source regions and form concentration gradients by undergoing diffusion and degradation. Morphogen gradients play an important role in body axis formation, growth control of tissues, and also branching morphogenesis \cite{stapornwongkul_2021,kicheva_2023}. For example, the mouse mammary gland produces transforming growth factor-$\beta$ (TGF-$\beta$) which forms gradients away from the gland and by its inhibitory effect on branch growth ensures the regular spacing of mammary gland branches \cite{sternlicht_2006,daniel_1996,daniel_1989,silberstein_1987,hannezo_2017}. 
Apart from morphogens, branching morphogenesis can also be guided by external guiding cues, e.g., mechanical constraints imposed by the surrounding musculature \cite{kim_2015,koser_2016,oliveri_2021,seo_2020,ucar_2021}. However, challenges in understanding branching morphogenesis include the specification of the often dynamic and transient morphogen source regions during branching morphogenesis. Here, the self-amplifying protrusions in physical branching systems provide an attractive possibility. 

A further interesting dimension of biological branching systems is their capacity for tremendous growth and remodeling, which may contribute to the emergence of allometric scaling laws \cite{peters_1983,schmidt-nielsen_1984}. Examples here include the expansion of the human lung surface area or the capacity of the vascular network during postembryonic growth or in response to physical exercise \cite{tenney_1963,nelson_1990,bloor_2005}. Also the branched planarian gut is subject to tremendous remodeling during regeneration or the food-supply dependent growth and degrowth of the animals \cite{forsthoefel_2011}, which amounts to more than $40$-fold variations in body length or more than $800$-fold variations in cell number in case of the model species \textit{S. mediterranea} \cite{thommen_2019}. Although it is known that the highly branched planarian gut and all other planarian organs qualitatively scale with body size, neither the quantitative scaling laws nor the branching morphogenesis mechanisms are currently understood. 

In this paper, we provide a theoretical framework to study the growth and scaling of branched organs. Our framework conceptualizes branching morphogenesis by the dynamics of a moving interface (e.g., organ outline). Specifically, we envisage the growth of the interface under the control of morphogens and study the resulting structures in a growing domain to represent organism growth. We use the scaling of the branching patterns of the planarian gut as biological inspiration and parametrize our model on the basis of quantified scaling behaviors. The fact that our model can recapitulate the self-organized emergence of gut-like branching patterns and their scaling behaviors demonstrates the utility of instability-driven morphogen-controlled interface growth in generating branched structures. Overall, our model integrates interface growth, morphogen dynamics, and organism growth and therefore provides a general framework for explaining the self-organized scaling of branched structures.

\begin{figure*}
    \centering
    \includegraphics[scale=0.9]{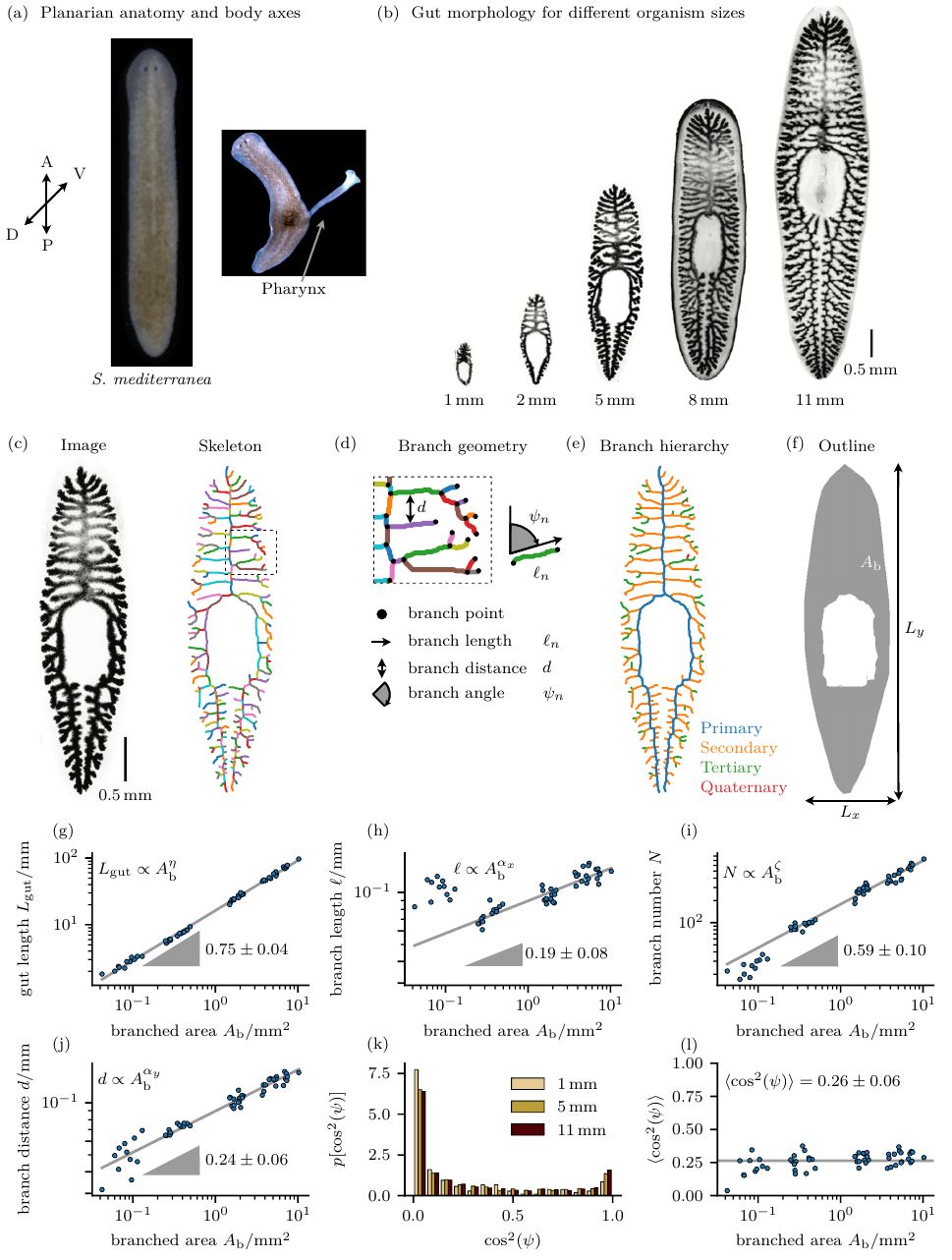}
    \caption{\textbf{Size-dependent gut properties in \textit{S. mediterranea}.} (a) Specimen of the model species \textit{S. mediterranea}. On their ventral side, planarians possess a muscular tube (pharynx) that is extended for food ingestion. Right most image is from \cite{rossant_2014}. (b) Representative whole-mount \textit{in-situ} hybridization specimens of the indicated sizes, labeled with a probe against the gut-specific transcript dd\_Smed\_v6\_75\_0\_1. Maximum projections of confocal z-stacks are shown. (c) We determine the gut skeleton, a one pixel wide, topology-preserving representation of the original image. (d) We identify branches (colored lines) as the set of pixels that connects branch points (black dots) and quantify branch length $\ell_n$, branch angle $\psi_n$, and branch distance $d$, where $n$ labels an individual branch. (e) Branches can be grouped into different hierarchy levels. (f) The convex hull of the skeleton (gray) serves as a measure for the organismal outline and allows us to determine organism width $L_x$, organism length $L_y$, and the branched area $A_{\textrm{b}}$. (g-j) Gut features (dots) as a function of branched area along with a power law fit (gray line) and the corresponding scaling exponent $\pm$ standard deviation. The power law fit was obtained from a fit of a linear function to the logarithmized data. Note that in (h) and (i) the fit was performed for organism sizes $A_{\textrm{b}}>\qty{0.2}{\milli\metre\squared}$ to avoid the biasing effect of the posterior part of the primary branch in small organism sizes. (k) Average orientational order parameter $\langle\cos^2(\psi)\rangle$ as a function of branched area. We additionally show the mean orientation order parameter across organism sizes (gray line). (l) Distribution of orientational order parameter for different organism sizes.
    \label{fig:quantification}}
\end{figure*}

\section{Morphometry and allometric scaling of the Planarian gut}\label{sec:morphometry}
\subsection{Quantification of planarian gut morphology}
We use the planarian gut as an experimental model system to study branched organ scaling (Fig.~\ref{fig:quantification}a). Their flat body is bilaterally symmetric with a distinct anterior-posterior (head and tail region) and dorso-ventral polarity (photoreceptors on dorsal side). On their ventral side, a muscular tube, the so-called pharynx, can extrude from the body. The pharynx serves as the only body opening of the organism and is responsible for both ingestion of food and excretion of waste.
The planarian gut spans the entire organism (Fig.~\ref{fig:quantification}b). The gut has a characteristic shape with one primary branch in the anterior and two primary branches in the posterior body half. Secondary side branches project laterally from the primary branches and branch again into higher-order side branches, thus giving rise to the highly branched morphology of the organ. 

To quantify gut branching morphologies, we determined the gut skeleton from binarized images of gut \textit{in-situ} hybridizations (Fig.~\ref{fig:quantification}c).
The skeleton is a one pixel wide, topology-preserving representation of the binarized image and allows us to quantify various aspects of gut branching morphologies. We determine individual branches in the skeleton and on the basis of this determine several geometrical branch properties (Fig.~\ref{fig:quantification}d, details see appendix~\ref{app:image_analysis}). We quantify the length $\ell_n$ of an individual branch, where $n=1,\dots,N$ with $N$ being the total number of branches. From this we determine the organismal mean branch length $\ell$. We quantify the total gut length $\Lgut$ as the sum of the individual branch lengths. Further, we also quantify branch distance $d$. Finally, we use the circular mean orientation $\psi_n$ of individual branch segments as a measure for branch orientation \cite{mardia_2000}.
The gut skeleton also allows us to determine branch hierarchy (Fig.~\ref{fig:quantification}e). We call the set of branches that form the longest path from the gut origin ``primary branch'' and similarly group the remaining branches according to their length. Knowing the branch hierarchy allows us to analyze hierarchy-dependent branch properties.
Finally, we also employ the skeleton to quantify organism size (Fig.~\ref{fig:quantification}f). To this end, we determine the convex hull of the skeleton and use the major and minor axis length of the convex hull as a measure for organism length $L_y$ and organism width $L_x$, respectively. Additionally, we determine branched area $\Ab=\Aw - \Ap$ from the difference of worm area $\Aw$ and pharynx area $\Ap$ as a measure for the area the gut can grow into.
Overall, we have established the quantification of key gut and organism properties, which allows us now to study size-dependent gut properties.

\subsection{Scaling of planarian gut morphology}
We next use our gut quantification to study the size-dependence of gut morphologies. We first study the size-dependence of the total gut length $\Lgut$ and find a strong increase of total gut length with branched area $\Ab$. Interestingly, the size-dependency of $\Lgut$ is well-described by the power law relationship $\Lgut \propto \Ab^{\eta}$ with scaling exponent $\eta=0.75\pm0.04$ (Fig.~\ref{fig:quantification}g). 
We next investigate the size-dependence of mean branch length $\ell$ and the total number of branches $N$ (Fig.~\ref{fig:quantification}h,i). We find that both $\ell$ and $N$ contribute to total gut length increase. Both quantities independently display a size increase that is approximated by the power law scaling $\ell\propto\Ab^{\alpha_x}$ with $\alpha_x=0.19\pm0.08$ and $N\propto\Ab^{\zeta}$ with $\zeta=0.59\pm0.10$, for large organism sizes ($\Ab>\qty{0.2}{\milli\metre\squared}$). For small organisms ($\Ab<\qty{0.2}{\milli\metre\squared}$), we observe a different trend as the two long posterior branches of the primary branch bias the mean branch length $\ell$ towards larger values.
We next study the size-dependence of branch distance and find that its size increase is well described by the power law $d\propto \Ab^{\alpha_y}$ with $\alpha_y=0.24\pm0.06$ (Fig.~\ref{fig:quantification}j). Note that the scaling of total gut length with branched area implicitly defines a length $\Ab/\Lgut$. This length serves as a measure for branch distance and scales with scaling exponent $0.25$ in agreement with the value found for $\alpha_y$. 
Finally, we study branch orientation and its size-dependence. To this end, we first consider the branch orientational order parameter distribution $\cos^2(\psi)$. We find a strong peak at $\cos^2(\psi)=0$ and $\cos^2(\psi)=1$ indicating branches with vertical and horizontal orientation, respectively (Fig.~\ref{fig:quantification}k). Additionally, we find that distribution shape is independent of organism size. The size-independent branch orientation is additionally supported by the size-independent mean orientational order parameter (Fig.~\ref{fig:quantification}l). Overall, we have therefore established a set of complementary scaling relationships that characterize the size-dependent properties of the planarian gut.

\section{Morphogen-controlled interface dynamics}\label{sec:model}

\begin{figure*}
    \centering
    \includegraphics{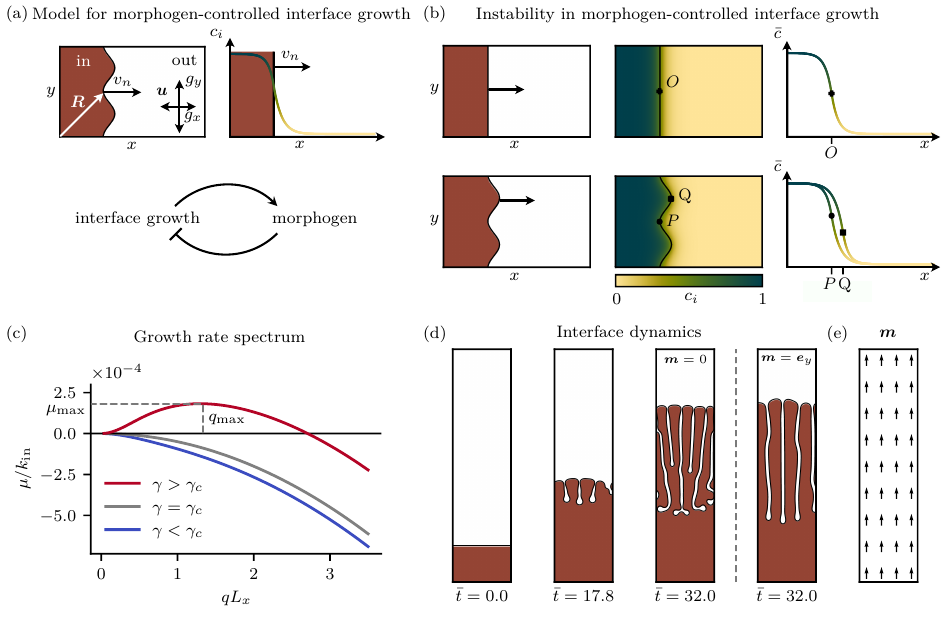}
        \caption{\textbf{Self-organized growth of branched morphologies from unstable interface growth.} (a) Left: In our model we represent the shape of an organ by a moving interface that separates the system into a region inside (``in'') and outside the organ (``out'') . Interface positions are described by the vector $\vecb{R}$ (black). The interface moves with normal velocity $v_n$ in a domain growing with rate $g_x$ and $g_y$ in the respective direction. Right: Morphogen is produced in the inside region and we find gradients of morphogen concentration $c_i$ towards the outside region, where $i=\textrm{in},\textrm{out}$. Bottom: We consider a mutual feedback between interface growth, which controls morphogen production, and morphogen concentration, which inhibits interface growth.
    (b) Top row: For a flat interface, interface motion is controlled only by the morphogen concentration at the interface position ($O$). Every point on the interface has the same morphogen concentration and thus every point of the interface experiences the same normal velocity.
    Bottom row: The motion of a curved interface depends on the morphogen concentration at the interface position and additionally the interface curvature $\kappa$.
    We find that the morphogen concentration at tips ($Q$) is reduced compared to valleys ($P$) and thus growth of tips is enhanced compared to growth of valleys. A positive feedback between tip growth and morphogen reduction at tips leads to unstable interface growth.
    (c) Growth rate $\mu$ of an interface perturbation with wavevector $q$ obtained from linear stability analysis. We additionally indicate the position of the maximal growth rate $q_{\textrm{max}}$ and the corresponding growth rate $\mu_{\textrm{max}}$ for different values of the inhibition strength $\gamma$ (color).
    (d) Formation of a branched pattern from an initially flat interface governed by Eq.~\eqref{eq:interface_motion} and \eqref{eq:morphogen_dynamics} without orientation field ($\vecb{m}=0$) and with orientation field ($\vecb{m}=\vecb{e}_y$).
    (e) Orientation field $\vecb{m}=\vecb{e}_y$.}
    \label{fig:model_illustration}
\end{figure*}

\subsection{Model definition}
We develop a minimal model for self-organized branching morphogenesis using a continuum theory in 2D (Fig.~\ref{fig:model_illustration}a). We represent the shape of a developing organ by the dynamics of a moving interface that separates the branching organ (region labeled ``in'') from the rest of the organism (region labeled ``out''). We describe positions of the interface by the vector $\vecb{R}$ with the local unit normal to the interface described by the vector $\vecb{n}$. Organ growth and therefore interface growth are controlled by a diffusing morphogen. We denote the morphogen concentration in each of the regions $i=\textrm{in},\textrm{out}$ by $c_i(\vecb{x},t)$, where $\vecb{x}=(x,y)$ denotes a position in the organism. Finally, we take into account tissue growth by a velocity field $\vecb{u}$ with $g=\vecb{\nabla}\cdot\vecb{u}$ as the area growth rate. 

\textit{Interface dynamics.}---The dynamics of the interface position $\vecb{R}(t)$ provides a coarse-grained description of organ shape development. During morphogenesis, the interface moves in normal direction $\vecb{n}$ according to
\begin{equation}\label{eq:interface_motion}
    \partial_t \vecb{R} = (v_n + u_n) \vecb{n}.
\end{equation}
Here, the interface velocity relative to the tissue $v_n$ and the contribution to interface velocity from organism growth $u_n$ are given by
\begin{subequations}\label{eq:interface_motion_details}
\begin{align}
    v_n &= \chi(c,\vecb{m}) - \beta \kappa \label{eq:interface_motion_details_vn}\\
    u_n &= \vecb{u}(\vecb{R})\cdot \vecb{n}.
\end{align}
\end{subequations}
The normal velocity $v_n$ is regulated by the morphogen concentration $c_i$ and an orientation field $\vecb{m}$ via the function $\chi$. Furthermore, interface motion depends on the local curvature $\kappa$ of the interface as described by the coefficient $\beta>0$. Such curvature dependencies arise naturally for example via surface tension effects. Here, a protrusion has positive curvature and therefore reduces interface velocity according to Eq.~\eqref{eq:interface_motion_details}. As a consequence, the curvature dependence of interface velocity stabilizes the interface.
Finally, we consider that interface motion takes place in a growing tissue with tissue velocity field $\vecb{u}$. Therefore  the interface experiences an additional advection $\vecb{u}\cdot\vecb{n}$.

We consider a minimal model that can generate branched morphologies similar to those observed in the gut. We choose $\chi(c,\vecb{m})$ of the product form
\begin{equation}\label{eq:bias_parameter}
		\chi(c,\vecb{m}) = \Gamma(c)\Theta(\vecb{m}),
\end{equation}
where $\Gamma(c)$ captures the regulation of interface velocity by morphogen concentration and $\Theta(\vecb{m})$ captures the influence of the orientation field $\vecb{m}$ on the interface velocity. We consider a linear dependence of interface velocity on morphogen concentration:
\begin{equation}\label{eq:morphogen_inhibition}
    \Gamma(c) = v_0 - \gamma c(\vecb{R})\,.
\end{equation}
Here, $v_0$ is the interface velocity in the absence of a morphogen and the coefficient $\gamma$ describes inhibition ($\gamma>0$) or activation ($\gamma<0$) of interface motion. The orientation field $\vecb{m}$ aligns interface velocity as described by the function
\begin{equation}\label{eq:angle_dependence}
    \Theta(\vecb{m}) = 1 - \delta (1-\vecb{n}\cdot\vecb{m}),
\end{equation}
where the coupling parameter $\delta$ with $\delta\in[0,1/2]$ controls the influence of the external orientation field on interface velocity. For $\delta=0$ there is no orientational bias, for $\delta>0$ interface growth is biased in the direction of the orientation field. For interface normal parallel to the orientation field ($\vecb{n}\cdot\vecb{m}=1$), the interface dynamics is maximal and controlled only by the morphogen concentration at the interface. However, for interface normal perpendicular to the orientation field ($\vecb{n}\cdot\vecb{m}=0$), interface motion is suppressed by a factor $1-\delta$.

\textit{Morphogen dynamics.}---The morphogen concentration is subject to the reaction-diffusion-advection equation
\begin{equation}\label{eq:morphogen_dynamics}
    \partial_t c_i +  \vecb{u}\cdot\vecb{\nabla} c_i= D \nabla^2 c_i - (k_i + \vecb{\nabla}\cdot \vecb{u}) c_i + s_i
\end{equation}
together with the boundary conditions at the interface $\vecb{R}$
\begin{subequations}
\label{eq:bc_morphogen}
    \begin{align}
        \cin(\vecb{R}) &= \cout(\vecb{R}) \\
        \vecb{n}\cdot\nabla\cin(\vecb{R}) &= \vecb{n}\cdot\nabla\cout(\vecb{R}).
    \end{align}
\end{subequations}
Morphogen diffuses with effective diffusion coefficient $D$ and is advected with tissue velocity $\vecb{u}$. It is produced at region-dependent rate $s_i$ and is degraded at region-dependent rate $k_i$. It is diluted at rate $\vecb{\nabla}\cdot \vecb{u}$ due to organism growth. The balance of production and degradation defines the steady-state concentration value $c^0_i=s_i/k_i$ in a homogeneous and non-growing tissue. Diffusion and degradation define the characteristic length scales $\lambda_i=\sqrt{D/k_i}$ in each region.
In the following we consider the case where $\Delta c = \cin^0 - \cout^0> 0$ corresponding to more morphogen production in the ``in'' than the ``out'' region.

\textit{Tissue growth.}---We consider homogeneous and anisotropic organism growth, where every point $\vecb{x}$ in the tissue moves according to
\begin{subequations}
\label{eq:organism_growth}
    \begin{align}
    \partial_t \vecb{x} &= \vecb{u}\\
    \vecb{u} &= g_x \vecb{e}_x + g_y \vecb{e}_y,
\end{align}
\end{subequations}
where $g_x$ and $g_y$ denote the tissue growth rate in $x$ and $y$ direction, respectively. Hence, the area growth rate obeys $g=g_x+g_y$.

Using this minimal model for morphogen-controlled organ morphogenesis and growth we can now study how branching morphologies can emerge via dynamic instabilities.

\subsection{Shape instability in morphogen-controlled interface growth}
We discuss the basic concepts that underlie the instability of the interface and provide a linear stability analysis. We then show examples of how this instability gives rise to branched interface patterns.

We consider a 2D rectangular domain of length $L_y$ and width $L_x$ in the absence of growth, $\vecb{u}=0$. We first study morphogen profiles in two simple interface configurations and large $L_x$.
The first configuration is a flat interface for which using Eq.~\eqref{eq:morphogen_dynamics} we find the corresponding steady-state morphogen concentration gradient which decays with increasing distance $x$ from the interface (Fig.~\ref{fig:model_illustration}b, top). Eq.~\eqref{eq:interface_motion_details_vn} provides the interface velocity which is constant along the interface.  
The second configuration is a periodically modulated interface. The corresponding morphogen gradient is different near a tip and a valley with lower concentration at tips (Q) than at valleys (P) (Fig.~\ref{fig:model_illustration}b, bottom). As a result, the interface velocity also differs at tips and valleys. The interface velocity is influenced by both curvature and morphogen concentration which represent competing influences on the interface. The curvature dependence decreases the velocity at the tip (Q) and increases it at the valley (P) and thereby stabilizes interface motion. In contrast, the morphogen increases motion at the tip and decreases it at the valley therefore having a destabilizing effect. Depending on which of the competing effects dominate the interface either stabilizes towards a flat interface (curvature effects dominate) or the interface destabilizes by advancing at the tips (morphogen inhibition dominates). In the latter case, morphogen inhibition provides positive feedback enabling the self-organized formation of a branched pattern.

To quantitatively study the emergence of patterns, we perform a linear stability analysis of interface motion. To this end, we consider a flat interface moving with constant velocity $v$ and determine the growth rate $\mu$ of sinusoidal interface perturbations with wavevector $q=2 \pi/\lambda$ and corresponding wavelength $\lambda$ (see appendix~\ref{sec:stability_analysis}). 
The growth rate as function of wave vector is shown in Fig.~\ref{fig:model_illustration}c for different values of the parameter $\gamma$ describing the inhibiting effect of the morphogen. 
For $\gamma$ larger than a critical value $\gamma_c$, $\mu$ becomes positive indicating that the system becomes unstable. In the unstable regime ($\gamma>\gamma_c$), the growth rate exhibits a maximum $\mu_{\textrm{max}}$ at wavevector $q_{\textrm{max}}$ with corresponding wavelength $\lambda_{\textrm{max}}$ indicating that perturbations with this wavelength grow fastest and therefore dominate the pattern formation at early stages. As a consequence, a pattern with wavelength $\lambda_{\textrm{max}}$ will form and we can use $\lambda_{\textrm{max}}$ as a measure for branch distance and $\mu_{\textrm{max}}$ as a measure for branch growth rate. By using the limit of small interface velocity ($v \ll D/\lambda_{\textrm{max}}$) and small branch distance ($\lambda_{\textrm{max}}\ll\lambda_i$) we find
\begin{equation}\label{eq:pattern_lengthscale}
    q^3_{\textrm{max}} \simeq \frac{\gamma \Delta c}{\beta}\frac{1}{\lambdain \lambdaout}
\end{equation}
as an approximation for the wavevector $q_{\textrm{max}}$. This relation shows the dependence of the characteristic length scale and therefore the branch distance on model parameters.

\subsection{Numerical solution}
To study the formation of interface patterns describing the branching organ shape, we use a phase field method to capture the interface dynamics (see appendix~\ref{app:phase_field_model}). We first study the dynamics of a flat interface in a rectangular, non-growing domain in the absence of an orientation field.
We choose parameters corresponding to the regime of unstable interface motion and find that the flat interface undergoes an instability with a characteristic length scale leading to the formation of several branches of increasing length (Fig.~\ref{fig:model_illustration}d). We additionally study interface motion subject to an orientation field ($\vecb{m}=\vecb{e}_y$) demonstrating that the orientation field $\vecb{m}$ can influence branch orientation and morphology (Fig.~\ref{fig:model_illustration}e).

\section{Branching dynamics in non-growing domains}\label{sec:morphogenesis_non_growing}

\begin{figure*}
    \centering
    \includegraphics{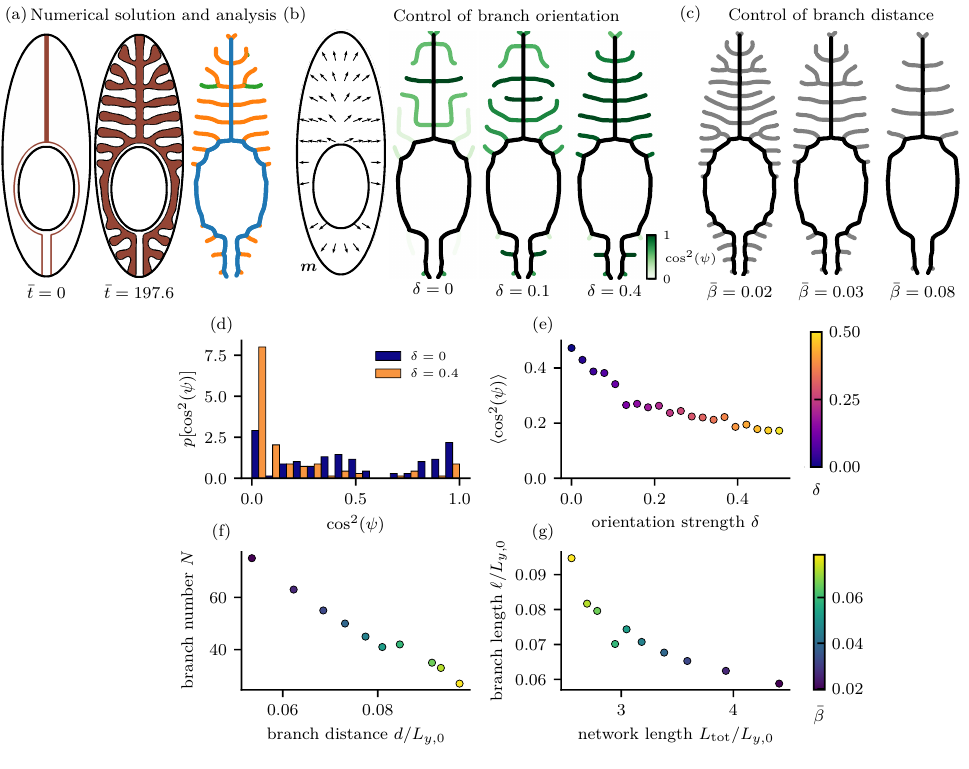}
    \caption{\textbf{Control of branched patterns in a non-growing domain.} (a) We show the boundary conditions employed in our model (left) along with a representative numerical solution (middle) and the corresponding skeleton (right). $\bar{t}=t\mumax$ indicates rescaled time. (b) We show the orientation field $\vecb{m}$ employed in our model along with skeletons of steady-state gut structures for different values of the orientation strength $\delta$. We highlight side branches (green) and the primary branch (black). For side branches, the color code indicates the orientational order parameter $\cos^2(\psi)$. (c)~Skeletons of steady-state gut structures for different values of the curvature dependence $\bar{\beta}=\beta/(\gamma \Delta c\lambdain)$. Side branches are indicated in gray. (d)~Orientational order parameter distribution $p[\cos^2(\psi)]$ for $\delta=0$ (blue) and $\delta=0.4$ (orange). (e)~Average orientational order parameter $\langle \cos^2(\psi)\rangle$ as a function of the orientation strength $\delta$. The average was obtained from $n=5$ samples. (f,g) Geometrical properties of branched patterns for different values of curvature dependence $\bar{\beta}$.}
    \label{fig:model_non_growing}
\end{figure*}

\subsection{Initial and boundary conditions motivated by planarian morphogenesis}
We now ask whether the interface dynamics can generate branched patterns similar to the morphology of the planarian gut.
We represent the shapes of the organism and the pharynx by an outer and inner ellipse, respectively, defining the model domain. We initialize the primary branch by a line of width $\bpb$ in the anterior region along the symmetry axis connected by two lines bordering the inner ellipse leading to two parallel lines in the posterior region (see brown line in Fig.~\ref{fig:model_non_growing}a). These lines provide the boundary conditions $\phi=1$ for the phase field that describes the organ shape. Additional boundary conditions are $\vecb{n} \cdot \vecb{\nabla} \phi=0$ at the domain boundaries. The initial condition for the concentration field is $c=0$ and boundary conditions are $\vecb{n} \cdot \vecb{\nabla} c=0$ at all domain boundaries.
Fig.~\ref{fig:model_non_growing}a shows an example of a solution of the dynamical equations Eq.~\eqref{eq:interface_motion} using these boundary conditions and taking into account an orientational cue. Shown is a resulting branched pattern at steady-state (middle) and the corresponding skeleton (right). This example is similar to the morphology of a branched organ of a small flatworm (see Fig.~1).

\subsection{Control of branch orientation}
We first study the influence of the orientation field $\vecb{m}$ on branch morphology. To this end, we study the interface dynamics varying the parameter $\delta$ describing the coupling of the interface to the orientation field (Fig.~\ref{fig:model_non_growing}b, right). For $\delta=0$, we find in the steady-state a morphology where branches have different orientations to the domain boundary. For example, some branches grow along the organism boundary, some are taking sharp turns. Increasing $\delta$, branches become more parallel to each other and tend to point towards the domain boundary.

The morphology of the branched patterns can be captured by the branch orientation statistics. Fig.~\ref{fig:model_non_growing}d shows the orientation distribution $p[\cos^2(\psi)]$ for two values of $\delta$, revealing that for larger values of $\delta$ the orientations become more aligned perpendicular to the long axis.
The alignment of orientation for increasing $\delta$ is reflected in the orientational order parameter $\langle \cos^2(\psi) \rangle$, which shows a decrease for increasing $\delta$ corresponding to branches increasingly pointing toward the domain boundary (Fig.~\ref{fig:model_non_growing}e). 
Overall, coupling to the orientation field $\vecb{m}$ controls branch orientation and therefore pattern morphology. 

\subsection{Control of branch distance}
A further key feature of network morphology is branch distance. Branch distance is related to the length scale $q_{\textrm{max}}$ that emerges at the interface shape instability which is given by Eq.~\eqref{eq:pattern_lengthscale}. The equation suggests that the branch distance can be regulated by the parameter $\bar{\beta}=\beta/(\gamma \Delta c \lambdain)$. Fig.~\ref{fig:model_non_growing}c shows steady-state patterns for different values of $\beta$, revealing that indeed the branch distance can be controlled.
As the branch distance is increased the branch number decreases correspondingly (Fig.~\ref{fig:model_non_growing}f). 
Furthermore, we show the mean branch length and the total pattern length showing that large networks contain many short branches (Fig.~\ref{fig:model_non_growing}g). Large networks contain more branch points and therefore consists of shorter branches.

So far we have discussed the formation of branched patterns in a non-growing domain. We now study the dynamics of branched patterns that form in a growing domain.

\section{Branching dynamics in growing domains}\label{sec:morphogenesis_growing}
\begin{figure*}
    \centering
    \includegraphics{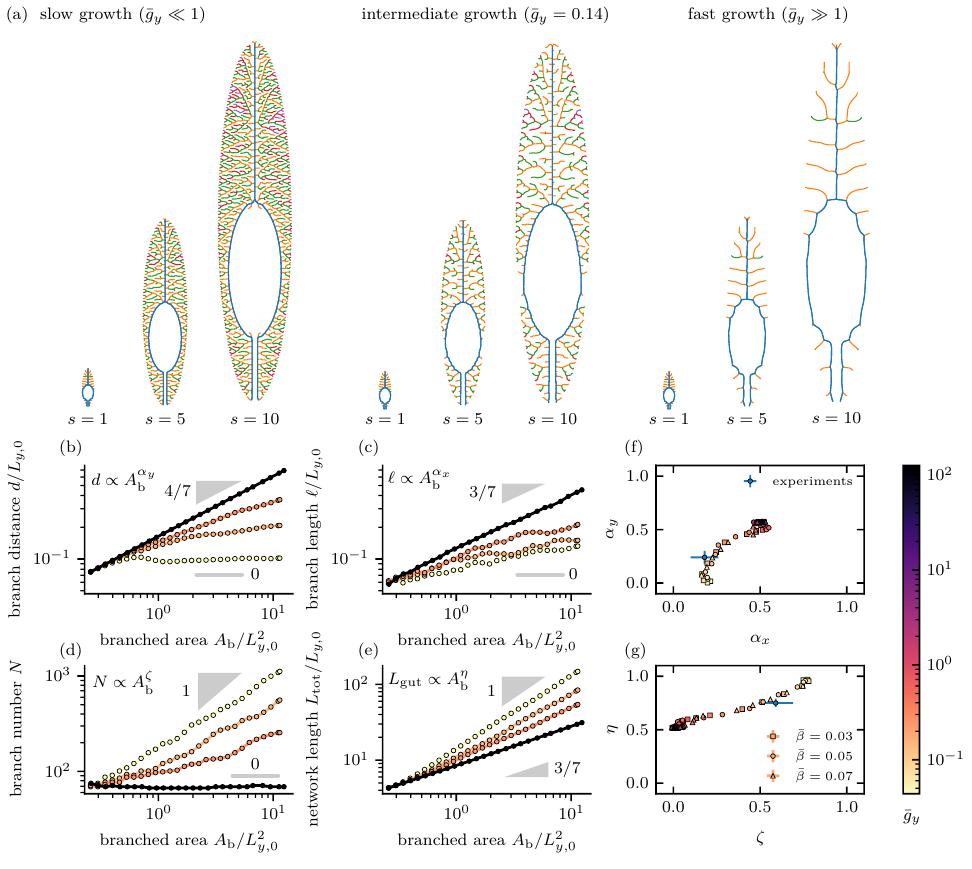}
    \caption{\textbf{Size-dependent properties of branched patterns in growing domains.} (a) Skeleton of simulated gut structures for different values of the relative size increase $s=L_{y}/L_{y,0}$ for different domain growth rates $\bar{g}_y=g_y/\mumax$. (b-e) Geometrical properties of branched patterns as a function of branched area $\Ab$. Additionally, we provide approximate scaling exponents for comparison (gray). (f,g) Scaling exponents for the features of branched patterns presented in (b-e). Scaling exponents were extracted from a linear fit of the logarithmized data for large system sizes ($\Ab/L_{y,0}^2>0.4$). We additionally show the corresponding scaling exponents in the planarian gut (blue dot).}
    \label{fig:model_growing}
\end{figure*}

\subsection{Influence of domain growth on branching dynamics}
To investigate the influence of domain growth on branched patterns, we use branched patterns obtained in a non-growing domain as an initial condition and study their time evolution using Eq.~\eqref{eq:interface_motion} and Eq.~\eqref{eq:morphogen_dynamics} taking into account the velocity field $\vecb{u}$ accounting for domain growth. We consider domain size increases by a factor $s=L_y/L_{y,0}$ while $L_x$ grows as $L_x\propto L^p_y$. Here, $p=g_x/g_y$ describes growth anisotropy with $g_x$ and $g_y$ denoting the growth rates along the long and short axis of the domain.
We choose $p=0.75$ as estimated from experimental data (see appendix Fig.~\ref{fig:length_area_worm}).
We find that the ratio $\bar{g}_y=g_y/\mumax$ of the domain growth rate $g_y$ and the branch growth rate $\mumax$ has a strong influence on pattern morphology (Fig.~\ref{fig:model_growing}a). For slow domain growth ($\bar{g}_y\ll1$), the number of branches increases to keep the branch distance roughly constant. New branches either form by tip splitting or by generation of new side branches along the primary branch and also higher order branches. In the limit of fast domain growth ($\bar{g}_y\gg1$), the branch distance increases with time as the pattern is essentially rescaled and no new branches form.
In the intermediate growth regime, new branches form while the branch distance on average increases.

\subsection{Size-dependent properties of branched patterns}
We next quantified the properties of branched patterns as a function of domain area $\Ab\propto L_y^{1+p}$ for different growth rates. We find that for small systems branch distance behaves independent of growth rate and scales as $d\propto L_y$. In this case the increase of branch distance is too small for new branches to form. For larger domain sizes, the branch distances then diverge for different growth rates at larger sizes as new branches can be formed (Fig.~\ref{fig:model_growing}b). The figure reveals that for slow growth rates the branch distance remains roughly constant for large domains while for fast growth rates it scales with domain size, $d\propto L_y$. 
For large domain sizes ($\Ab/L^2_{y,0}\gg1$), the competition between branch formation and domain growth results in a scaling of branch distance $d\propto \Ab^{\alpha_y}$with a growth rate dependent scaling exponent $0\leq\alpha_y\leq1/(1+p)$.

Branch length $\ell$ and branch number $N$ also follow growth rate dependent scaling laws as a function of domain area with scaling exponents $\alpha_x$ and $\zeta$ (Fig.~\ref{fig:model_growing}c,d). For fast domain growth ($\bar{g}_y\gg1$), branched patterns are scaled up without any morphological changes and as a consequence we find $\ell\propto L_x$ or $\alpha_x = p/(1+p)$ and $N$ remaining constant. By contrast, for small organism growth rates ($\bar{g}_y\ll1$), the scaling of branched patterns is characterized by the formation of side branches at constant branch distance. Therefore we find only a small increase in mean branch length, but a large increase in the total number of branches.
Finally, the size-dependent properties of total network length can be understood by making use of the relation $L_{\textrm{tot}}=N \ell$. For fast domain growth ($\bar{g}_y\gg1$), the total number of branches is constant and therefore the increase in total network length is due to an increase in mean branch length. By contrast, for slow domain growth ($\bar{g}_y\ll1$), the increase in total network length comes mainly from the increase in the number of branches (Fig.~\ref{fig:model_growing}e).
Overall, we observe power law scaling of various properties of branched patterns during domain growth. We find a range of scaling exponents controlled by the interplay of domain and branch growth with slow and fast domain growth rate as limiting regimes. We summarize our results by showing the range of scaling exponents from simulations in a parametric plot as a function of domain growth rate (Fig.~\ref{fig:model_growing}f,g).

Planarians grow if they are well fed but they can also shrink their body size under starvation. Motivated by this fact, we also studied properties of branched patterns subject to domain degrowth. In our model, branched network adjust their morphology to a shrinking domain, see appendix~\ref{app:periodic_degrowth}. Combining growth and degrowth, we find that branched patterns are maintained also for several periods of growth and degrowth (Fig.~\ref{fig:periodic_degrowth}).

Based on the results of this section, we next provide a detailed comparison between properties of branched patterns found in our model and in experiments.


\section{Comparison of the model to planarian gut morphology}\label{sec:comparison}
So far, we have systematically studied how model parameters influence branched pattern morphologies. We next compare the model to the planarian gut morphology choosing parameters for which realistic branched patterns emerge in the model. We adjust $\beta$ such that total network length matches the network length observed for small worms in the experiment. For larger domain sizes the total network length depends also on growth rate. We adjust the growth rate such that total network length matches the experimental data for both small and large worms. Furthermore, we choose a value of $\delta=0.4$ such that the value of the orientational order parameter $\langle \cos^2(\psi) \rangle$ shown in Fig.~\ref{fig:model_non_growing} matches the experimentally observed $\cos^2(\psi)\approx0.26$, see Fig.~\ref{fig:quantification}.

\begin{figure*}
    \centering
    \includegraphics{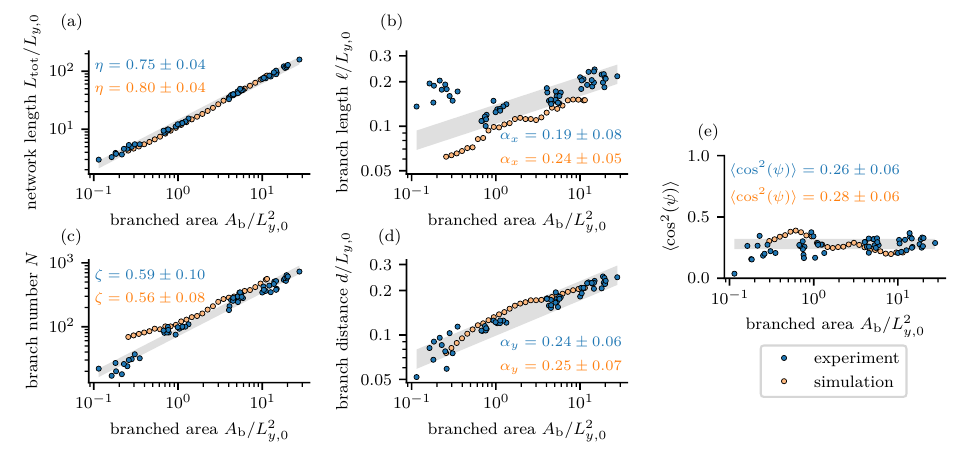}
    \caption{\textbf{Comparison of gut features in experiment and model.} We show gut features obtained in experiments along with features of branched patterns from our model by using the minimal system size $L_{y,0}$ as a characteristic length scale. The gray area corresponds to $\pm 15 \%$ of the power law fit to the respective feature of the planarian gut.}
    \label{fig:comparison_model_experiment}
\end{figure*}

In Fig.~\ref{fig:comparison_model_experiment}, we show the quantified properties of branched patterns obtained from numerical solutions of the model together with values measured in the planarian gut. We find good agreement between model and experiments for the key parameters quantified. These include total network length, branch distance, total number of branches, mean branch length and mean orientational order parameter. This comparison demonstrates that by adjusting only three parameters, our model can capture the size-dependence of five gut properties simultaneously. 

\section{Scaling theory for branching and growing networks}\label{sec:scaling_theory}
\subsection{Minimal network model}
\begin{figure*}
    \centering
    \includegraphics{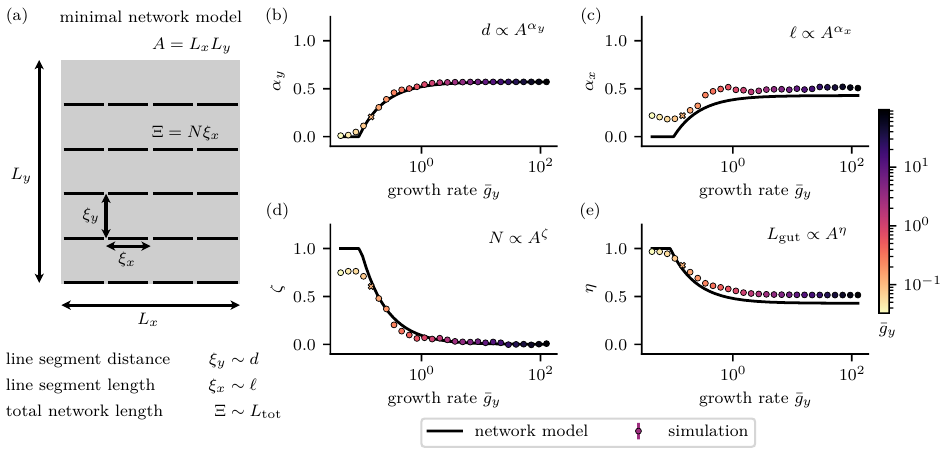}
    \caption{\textbf{Scaling theory for branching and growing networks.} (a) We reduce the complex morphology of branched patterns obtained with our model to a minimal network of straight line segments. Line segments of length $\xi_x$ are placed horizontally a distance $\xi_y$ apart in a rectangular system of width $L_x$, length $L_y$, and corresponding area $A$. (b) We show the scaling exponents obtained from branched patterns of our model (dots) along with the prediction from our scaling theory (line) for different values of growth rate $\bar{g}_y=g_y/\mu_{\textrm{max}}$ (color). To show the prediction from the scaling theory, we used the parameter values $r_x/\mumax=0.076$ and $r_y/\mumax=0.089$. The parameter values were obtained by manually matching Eq.~\eqref{eq:alphacase1} and Eq.~\eqref{eq:alphacase2} to the exponents $\alpha_x$ and $\alpha_y$ obtained from simulations.}
    \label{fig:scaling_argument}
\end{figure*}

We have shown that the scaling behavior of network properties depends on the ratio of the domain growth rate and the rate associated with the shape instability of branches. In order to better understand how scaling relations originate from the interplay of branch growth and organism growth, we now present a scaling argument using a minimal network model. We represent  branched patterns by a reduced network geometry, as shown in Fig.~\ref{fig:scaling_argument}a. Branches are represented by horizontal line segments of length $\xi_x$, spaced at a distance $\xi_y$, that constitute a network with $N$ line segments and total network length $\Xi=N\xi_x$. The quantities $\xi_x$, $\xi_y$, $\Xi$ correspond to quantities $\ell$, $d$, $L_{\textrm{tot}}$ measured in experiments and the continuum model. We use a rectangular geometry with area $A=L_x L_y$, where $L_x$ and $L_y$ denote the system width and length, respectively.

Considering a space-filling regular arrangement, we have $A= N \xi_x \xi_y$ and thus the total number of branches can be expressed as $N=A/(\xi_x \xi_y)$. Accordingly, the total network length is $\Xi=A/\xi_y$. The scaling exponents are defined as $\xi_x \propto A^{\alpha_x}$, $\xi_y \propto A^{\alpha_y}$, $N\propto A^{\zeta}$, and $\Xi\propto A^{\eta}$. From the definitions of network quantities we find the following relations between scaling exponents
\begin{subequations}\label{eq:network_exponents}
    \begin{align}
        \zeta &= 1 - \alpha_x - \alpha_y\\
    \eta &= 1 - \alpha_y.
    \end{align}
\end{subequations}

We consider the time evolution of the network in a domain growing at rates $g_x$ and $g_y$ along the $x-$ and $y-$direction. The line segment sizes $\xi_k$ and the system size $L_k$ with $k=x,y$ obey the dynamics
\begin{subequations}\label{eq:line_segment_dynamics}
    \begin{align}
        \frac{d L_k}{dt} &= g_k L_k \\
        \frac{d \xi_k}{dt} &= g_k \xi_k + r_k (\hat{\xi}_k - \xi_k), 
    \end{align}
\end{subequations}
where lengths are stretched according to domain growth. In the absence of growth the network relaxes to $\xi_k=\hat{\xi}_k$ at rates $r_k$, where $\hat{\xi}_k$ are characteristic length scales in the $x-$ and $y-$directions. These length scales relate to the branch width $\hat{d}$ and the branch length $\hat{\ell}$ defined by the patterning process in non-growing domain discussed in Section \ref{sec:morphogenesis_non_growing}, such that $\hat{\xi}_x\sim \hat{\ell}$ and $\hat{\xi}_y \sim \hat{d}$.  Eq.~\eqref{eq:line_segment_dynamics} describes a dynamics where line segment size is rescaled at rate $g_k$ due to domain growth, and line segment size relaxes to a preferred size $\hat{\xi}_k$ at rate $r_k$ due to the formation of new line segments, capturing the interplay between domain growth and network patterning. This dynamics is complemented by the initial conditions $L_k(t=0)=L_{k,0}$ and therefore $A(t=0)=L_{x,0}L_{y,0}$ and $\xi_k(t=0)=\xi_{k,0}$. 

\subsection{Emergence of scaling exponents}
Solving Eq.~\eqref{eq:line_segment_dynamics} for constant $g_k$, the system size increases exponentially according to $A(t)=A_0 e^{(g_x + g_y)t}$ and the growth of $\xi_k$ can be expressed for $g_k\neq r_k$ as
\begin{align}
	\xi_k(A) = \frac{\hat{\xi}_k}{1-\frac{g_k}{r_k}} + \left(\frac{A}{A_0}\right)^{\frac{g_k - r_k}{g_x + g_y}} \left[\xi_{k,0} - \frac{\hat{\xi}_k}{1-\frac{g_k}{r_k}}\right]. \label{eq:line_segment_scaling}
\end{align}
This equation allows us to identify regimes in which $\xi_k$ displays scaling with area. For large domain sizes ($A/A_0\gg1$) the ratio $g_k/r_k$ determines scaling behavior.
For slow domain growth, $g_k<r_k$, line segments have enough time to relax to their preferred distance and length. In this case, $\xi_k$ approaches a constant for large $A$, corresponding to 
\begin{equation}
    \alpha_k=0, \quad (g_k<r_k).\label{eq:alphacase1}
\end{equation}
For fast domain growth, $g_k>r_k$, line segment sizes increase with the domain size as $\xi_k \propto A^{\alpha_k}$ with 
\begin{equation}
    \alpha_k=\frac{g_k - r_k}{g_x + g_y}, \quad (g_k>r_k). \label{eq:alphacase2}
\end{equation}
In the special case $g_k=r_k$, the solution to Eq.~\eqref{eq:line_segment_dynamics} has the form 
\begin{equation}
    \xi_k(A)\propto \log(A/A_0), \quad (g_k=r_k). \label{eq:alphacaselog}
\end{equation}
We can also express the exponents $\eta$ and $\zeta$ using Eq.~\eqref{eq:network_exponents}, see appendix~\ref{app:network_exponents}. 

\subsection{Scaling theory of branched patterns}
We next apply our scaling argument to the scaling determined from numerical solutions of our continuum model. Fig.~\ref{fig:scaling_argument}b-e show scaling exponents $\alpha_x$, $\alpha_y$, $\zeta$, $\eta$ determined as a function of growth rate $g_y=(4/3) g_x$ (circles) together with the results of the simple network model (solid line). For large $g_y$, the exponents reach limits that are well described by the scaling theory: $\alpha_y\rightarrow g_y/(g_x+g_y)=4/7$,  $\alpha_y\rightarrow g_x/(g_x+g_y)=3/7$, $\eta\rightarrow g_x/(g_x+g_y)=3/7$, and $\zeta\rightarrow 0$. Parameters $r_x$ and $r_y$ have been determined from the data shown in Fig.~\ref{fig:scaling_argument}b,c using Eq.~\eqref{eq:alphacase2}. Overall, the  scaling behavior is in good agreement with the scaling exponents obtained from numerical solutions. They capture the behavior of four network properties as they change from small to large domain growth rates. The small discrepancies between minimal network model and continuum model could be due to simplifications of the geometry used in the minimal network model. 

\section{Discussion}
In this work, we have provided a theoretical framework to study branching morphogenesis in a growing tissue. In our framework, we represent a branched organ outline by an interface and study the morphogen-controlled dynamics of this interface in a growing domain. 
In particular, we show how an instability in interface dynamics leads to self-organized formation of branched patterns. Our model is motivated by the branching morphogenesis of the planarian gut which exhibits scaling as a function of organism size. With a minimal number of assumptions our model can reproduce key features of gut morphology including power laws of several geometric network properties. More generally, simple scaling arguments allow us to understand how the scaling behavior of a branching network emerges from the interplay between branch formation and domain growth.

Our work has parallels with, but also complements and extends previous approaches from non-equilibrium physics for studying the formation of branched morphologies such as solidification, viscous fingering, dielectric breakdown, or colony formation of bacteria \cite{meakin_1998,godreche_1991,vicsek_1992,langer_1980,ben_jacob_2000}.
These phenomena have in common with our approach that branched patterns emerge from a shape instability and that they can be oriented by guiding cues. For example, crystal growth can emerge via the Mullins-Sekerka instability and crystal axes can provide guiding cues \cite{mullins_1963,ohta_1988}. Viscous fingering occurs via the Saffman-Taylor instability and surface patterns can provide guiding cues \cite{saffman_1958,chen_1987}.
In contrast to these physical systems, in our approach a morphogen field governed by diffusion and degradation sets a characteristic length scale compared to a diffusive field giving rise to self-similarity. Furthermore, we study pattern formation in a growing domain and show that the interplay between interface dynamics and domain growth can give rise to power law scaling even for morphogen gradients with a characteristic length scale.

Previous work has studied self-organized growth using coarse-grained models based on rules for tip growth, branching and termination \cite{gavrilchenko_2024,hannezo_2017,bordeu_2023}. Such models can account for realistic branching morphology and provide a powerful framework to study network properties. By contrast, our approach provides a physical model for branch formation and dynamics based on interfacial instabilities and rules for tip dynamics naturally emerge. Our model is also coarse-grained and does not specify microscopic details, but it provides a physical scenario for branching morphogenesis.
Scaling behavior of branched networks has been discussed in previous work in the context of allometric scaling via transport network optimization \cite{west_1997,banavar_1999,bohn_2007,durand_2007}. Our work, which reveals how scaling properties could emerge, can provide a dynamical picture of how such optimized network architecture might be generated during development.

The biological inspiration for our model is the highly branched gut of planarians and the tremendous scaling this organ undergoes during the food-dependent growth and degrowth of these animals. The cellular and molecular mechanisms mediating planarian gut branching are generally poorly understood at the moment. For example, no secreted signaling molecules analogous to the morphogen in our model have so far been identified and the small number of genes known to influence gut branching are enriched for transcription factors and components of the actin cytoskeleton \cite{forsthoefel_2012,flores_2016,molina_2023}. The fact that our model reproduces the scaling behaviour of four different gut properties, as detailed above, makes the involvement of morphogen-dependent branch initiation a compelling motivation for the experimental identification of the biological mechanisms. Towards this goal, the effects of parameter changes on the gut pattern in our model provide helpful experimental starting points. For example, partial or complete loss of the morphogen in our model reduces the lateral inhibition between gut branches and results in increased branch thickness or fusion between lateral branches into a leaf-like morphology (see appendix Fig.~\ref{fig:morphogen_inhibition}). Similarly, the loss of the orientation cue from the boundary results in reduced alignment between secondary branches. Both should represent easily recognizable phenotypes in high-throughput screening experiments that are feasible in the planarian model system \cite{ivankovic_2019, reddien_2005}.

Another interesting interface between model and experiment are the dynamics of gut branching during growth and degrowth. In our model, lateral branching is often initiated at the tip of existing branches via tip splitting, due to the dilution of morphogen at the branch ends (see appendix Fig.~\ref{fig:branch_formation}). Interestingly, tip splitting and subsequent ``unzipping'' of the bifurcation has already been observed in pioneering live imaging studies of planarian gut branching \cite{forsthoefel_2011}, which we could also confirm as the predominant mode of branch addition in the anterior part of the animals (see appendix Fig.~\ref{fig:branch_formation}). Another interesting feature of our model is that under the specific parameter regime that recapitulates pattern scaling, branching patterns differ between growth and degrowth. Specific predictions include different gut properties at identical organism size depending whether the organism undergoes growth or degrowth (see appendix Fig.~\ref{fig:periodic_degrowth}).
Overall, the emergence of scaling during growth and degrowth presents fascinating challenges, both in terms of theory and biological mechanisms. As we have shown, the dynamic growth and degrowth of planarians provides a unique model system for exploring scaling by theory and experiment.

In essence, our model provides a general framework for the self-organized growth and scaling of branched networks. Our model can generate a range of scaling exponents depending on the ratio of branch growth rates and organism growth rates. The evolutionary significance of the particular values of scaling exponents that we measure in planarians is currently unclear. 
However, the central role of the gut as exchange surface for nutrients and the previous body of theoretical work on the putative relation between the structure of branching transport networks and metabolic scaling \cite{west_1997,banavar_1999} provides an intriguing direction for future exploration. Across animal phylogeny, metabolic rate scales as $P\propto M^{3/4}$ with animal mass $M$, which is known as Kleiber's law \cite{kleiber_1932}.
Although the mechanistic origins of the 3/4 exponent remain unknown, we have recently shown that planarian growth and degrowth follows Kleiber's law \cite{thommen_2019}.
Given the presence of the $3/4$ scaling exponent in both organ scaling ($\Lgut\propto \Ab^{3/4}$) and metabolic scaling ($P\propto M^{3/4}$), naturally the question regarding their relation arises. Our framework provides novel directions to probe possible relationships between branching morphogenesis and metabolic scaling. Such a relationship could provide new angles to understand the mechanistic basis of metabolic scaling. 

\begin{acknowledgments}
We would like to thank Sagnik Garai for a critical reading of the manuscript. We would like to thank Rick Kluiver, Jens Krull, and MPI-NAT animal services staff for worm care support. We acknowledge support from the Max Planck Computing and Data Facility.
\end{acknowledgments}

\bibliography{bibliography}
\newpage
\appendix
\section{Image analysis details}\label{app:image_analysis}
To analyze geometrical branch properties for gut structures obtained in simulations and experiments, we make use of the so-called skeleton. The skeleton is a one-pixel wide representation of the binarized simulation or experimental data with the same connectivity as the original data. Individual lattice site values indicate the presence ($n_{ij}=1$) or absence ($n_{ij}=0$) of gut structure.

\begin{figure}
    \centering
    \includegraphics{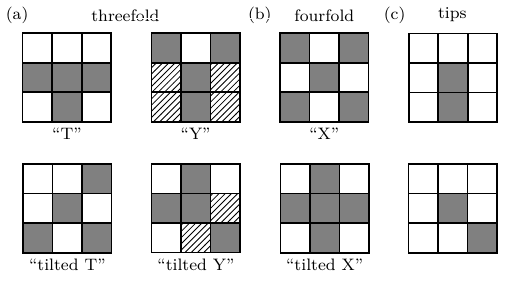}
    \caption{\textbf{Definition of branch points and tips.} The definition of branch points and tips is based on local $3\times3$ neighborhoods. (a) Representative neighborhoods that define threefold branch points. (b) Neighborhoods that define fourfold branch points. (c) Representative neighborhoods that define tips. Note that for the cases (a) and (c) also rotations by $90^{\circ}$ define the respective property.}
    \label{fig:vertex_definition}
\end{figure}

To identify individual branches in the skeleton, we first identify branch points based on local $3\times3$ neighborhoods in the skeleton (Fig.~\ref{fig:vertex_definition}). We call an occupied lattice site an $n$-fold branch point if the lattice size has $n$ occupied neighbors and a neighborhood with $n$ unoccupied non-connected regions. We call an occupied lattice site a tip, if it has exactly one connected neighbor. Finally, we define a branch $e_n$ as the set of pixels
\begin{equation}
	e_n = \{(i_{n,1},j_{n,1}),\dots,(i_{n,M},j_{n,M})\}\label{eq:definition_edges}
\end{equation}
that connects two branch points $(i_{n,1},j_{n,1})$ and $(i_{n,M},j_{n,M})$, where $M$ denotes the number of pixels that constitute the branch and $n$ is used to label individual branches. 

\subsection{Branch length, distance, and thickness}
By making use of the definition of a branch in Eq.~\ref{eq:definition_edges}, we can determine several geometrical branch properties. We define the branch length $\ell_n$ as the sum
\begin{equation}
	\ell_n = \sum_{m=1}^{M-1} \Delta \ell_{n,m},\label{eq:definition_branch_length}
\end{equation}
where $\Delta \ell_{n,m} = \sqrt{(i_{n,m+1} - i_{n,m})^2 + (j_{n,m+1} - j_{n,m})^2}$ denotes the Euclidean distance between two consecutive branch elements. We define the mean branch length as the arithmetic mean of individual branch lengths $\ell_n$ according to 
\begin{equation}
	\ell = \frac{1}{N} \sum_{n=1}^{N} \ell_n.\label{eq:definition_mean_branch_length}
\end{equation}
Finally, we define total gut length $\Lgut$ as the sum of individual branch lengths $\ell_n$
\begin{equation}
    \Lgut = \sum_{n=1}^{N} \ell_n\label{eq:definition_total_gut_length}
\end{equation}
from which follows that $\Lgut = N \ell$.

To determine branch distance, we consider the image columnwise and determine the vertical distance $d_k$ of sidebranches. Branch distance is then defined as the arithmetic mean of individual vertical branch distances according to
\begin{equation}
	d = \frac{1}{N_d}\sum_{k=1}^{N_d} d_k,\label{eq:definition_branch_distance}
\end{equation}
where $N_d$ denotes the total number of vertical distances considered.

\subsection{Branch orientation}
We use the mean branch angle $\psi_n$ to quantify branch orientation. The mean branch angle $\psi_n$ is defined as the circular mean of individual branch segment angles $\psi_{nm}$ \cite{mardia_2000}. To determine the circular mean of individual branch segment angles, we first determine the position $(x_{nm},y_{nm})$ on the unit circle corresponding to the angle $\psi_{nm}$ by
\begin{equation}
\begin{aligned}
 x_{nm} &= \cos \psi_{nm} \qquad\qquad&  y_{nm} &= \sin \psi_{nm} \\
        &= \frac{i_{n,m+1} - i_{n,m}}{\Delta \ell_{n,m}}             &  &=\frac{j_{n,m+1} - j_{n,m}}{\Delta \ell_{n,m}}
\end{aligned}
\end{equation}
and then determine the arithmetic average $(\bar{x}_n,\bar{y}_n)$ of the position on the unit circle. Finally, the average branched angle is defined as 
\begin{align}
	\tan(\psi_n) &= \frac{\bar{y}_n}{\bar{x}_n}.\label{eq:branch_orientation}
\end{align}
and can be obtained as $\psi_n=\textrm{atan2}(\bar{y}_n,\bar{x}_n)$, where $\textrm{atan2}$ denotes the two argument arctan. 
Using the circular mean instead of the arithmetic mean of individual branch angles allows us to also quantify the branch orientation of the curved branches in the gut.

\begin{figure}[b]
    \centering
    \includegraphics{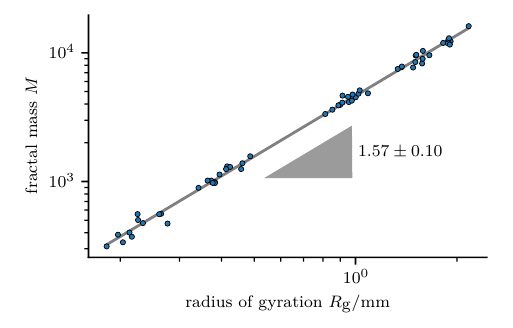}
    \caption{\textbf{Fractal properties of gut skeletons in experiments.} We show the fractal mass $M$ as a function of the skeleton radius of gyration $R_{\textrm g}$ defined in Eq.~\eqref{eq:rg}.}
    \label{fig:fractal_analysis}
\end{figure}

\section{Fractal properties of the planarian gut}
\begin{figure}
    \centering
    \includegraphics{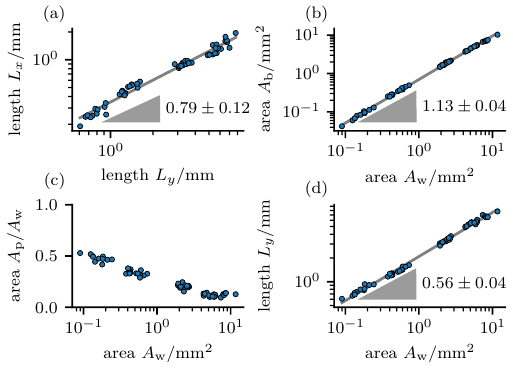}
    \caption{\textbf{Quantification of body size and aspect ratio of \textit{S. mediterranea}.} (a) Worm width $L_x$ as a function of worm length $L_y$ together with a fit to the power law $y=a x^b$ (gray line). (b) Branched area $A_{\textrm{b}}$ as a function of worm area $A_{\textrm{w}}$ together with a fit to the power law $y=a x^b$ (gray line). (c) Relative pharynx area $A_{\textrm{p}}/A_{\textrm{w}}$ as a function of worm area $A_{\textrm{w}}$. (d) Worm length $L_y$ as a function of worm area $A_{\textrm{w}}$ together with a fit to the power law $y=a x^b$ (gray line). We find $a=2.08\pm0.05$ and $b=0.56\pm0.04$.}
    \label{fig:length_area_worm}
\end{figure}
In the main text we discussed how size-dependent gut properties are captured by power law relationships. In particular, we demonstrated how the scaling of total gut length $\Lgut$ is captured by the scaling relationship $\Lgut\propto \Ab^{\eta}$ with scaling exponent $\eta$. Here, we now show how the allometric scaling of total gut length with branched area is further related to the fractal structure of the gut.

To quantify the fractal nature of the gut, we determine the fractal dimension $\df$ from the scaling $M\propto R_g^{\df}$ of fractal mass $M$ with the radius of gyration $R_g$ \cite{vicsek_1992}. We determine the fractal mass from the gut skeletons as $M = \sum_{ij} n_{ij}$ and the radius of gyration from
\begin{equation}
    R^2_{\textrm{g}} = \frac{1}{M} \sum_{i,j} (\vecb{r}_{ij} - \vecb{r}_{\textrm{cm}})^2, \label{eq:rg}
\end{equation}
where the vector $\vecb{r}_{ij}$ points to activated pixels. We find $\df=1.57\pm0.10$ indicating the fractal nature of the gut (Fig.~\ref{fig:fractal_analysis}).
To understand how the fractal dimension $\df$ is related to the scaling of total gut length and in particular the scaling exponent $\eta$, we next relate total gut length with the total mass $M$ and branched area with the radius of gyration. We use total gut length as an approximate measure for fractal mass ($\Lgut \propto M$) and branched area as an approximate measure for squared radius of gyration ($\Ab \propto R_g^2$). From this, we can easily determine that fractal dimension $\df$ and scaling exponent $\eta$ are related by $\df\simeq2\eta$ in good agreement with our earlier finding for $\df$.

\section{Allometric scaling of planarian body size}
We study different measures of organism size and their relation. We first study the relation between worm width $L_x$ and worm length $L_y$. We find the scaling relation $L_x\propto L_y^{0.79}$ (Fig.~\ref{fig:length_area_worm}a), where the scaling exponent $<1$ indicates that worm length increases faster than worm width. The scaling of worm width with worm length further allows us to determine the anisotropy parameter $p$. Given a constant organism growth rate $g_i$ in each direction, organism length and width are described by $L_i=L_{i,0}e^{g_i t}$. Their relation is given by $L_x\propto L_y^{g_x/g_y}$ and we find that the anisotropy parameter is given by $p=g_x/g_y$. We can determine the anisotropy parameter from the scaling of organism width and length and find  $p\simeq3/4$.
Next, we studied the relation between branched area $\Ab$ and worm area $\Ab$. We find the relation $\Ab\propto\Aw^{1.13}$ (Fig.~\ref{fig:length_area_worm}b), where the scaling exponent $>1$ indicates that branched area increases faster than worm area. which we can demonstrate independently (Fig.~\ref{fig:length_area_worm}c). Finally, we study the relation between worm length $L_y$ and worm area $\Aw$. We obtain the scaling relation $L_y\propto\Aw^{0.56}$ with a scaling exponent close to one as suggested by dimensional considerations. We note that our scaling relation for worm length and worm area is in agreement with the result $L_y\propto \Aw^{0.55}$ obtained independently in \cite{thommen_2019}.

\section{Phase field method for studying interface motion}
\label{app:phase_field_model}
\subsection{An overview of the phase field method}
To simulate the interface dynamics, we employ the phase field method \cite{kobayashi_2010}. In the phase field method the phase field parameter $\phi\in[0,1]$ is a central quantity that is used to implicitly define an interface. While $\phi=0$ denotes the ``in'' region, $\phi=1$ denotes the ``out'' region. Both regions are connected by thin transition layer of small, but non-zero width. The interface is then implicitly represented as the region between 
the ``in'' and the ``out'' region.

The dynamics of the phase field $\phi$ is defined on the basis of a Ginzburg-Landau-like energy $E$ of the form
\begin{equation}
	E[\phi] = \int d {\bm x} \left[\frac{K}{2}({\bm \nabla} \phi)^2 + f(\phi)\right].\label{eq:ginzburg_landau_energy}
\end{equation}
The first term introduces an energetic cost to form an interface and the second term is the energy density $f$ that denotes the contribution of the bulk to total energy. The energy density $f$ is composed of a symmetric part $f_{\textrm{s}}$ and a tilting part $f_{\textrm{t}}$ and is of the form
\begin{equation}
	f(\phi) = \frac{a^2}{2} f_{\textrm{s}}(\phi) + \frac{\hat{\chi}}{6} \; f_{\textrm{t}}(\phi)\label{eq:energy_bulk_part}
\end{equation}
Here, the parameter $a$ and the bias parameter $\hat{\chi}$ adjust the contribution of the symmetric and tilting part and therefore allow us to adjust the shape of the energy density. The symmetric part $f_{\textrm{s}}$ and tilting part $f_{\textrm{t}}$ are defined as
\begin{align}
	f_{\textrm{s}}(\phi)  &= \phi^2(1-\phi)^2   &f_{\textrm{t}}(\phi) &= \phi^2(2\phi-3).
\end{align}
The symmetric part is a double well potential with minima $\phi=0,1$ and therefore has $f_{\textrm{s}}(0)=f_{\textrm{s}}(1)=0$. The tilting part also obeys $f_{\textrm{t}}(0)=0$, but in contrast to the symmetric part has $f_{\textrm{t}}(1)=-1$. Overall, we then find $f(0)=0$ and $f(1)=-\hat{\chi}/6$. The parameter $\hat{\chi}$ allows us to control by how much $\phi=1$ is energetically preferred over $\phi=0$ and therefore implicitly controls interface motion. Note that the bias parameter underlies the restriction $|\hat{\chi}|<a^2$. For $|\hat{\chi}|<a^2$, Eq.~\eqref{eq:app_phase_field_equation_comoving} has stable stationary states at $\phi=0,1$. For $|\hat{\chi}|>a^2$, Eq.~\eqref{eq:app_phase_field_equation_comoving} has unstable stationary states at $\phi=0,1$.

The dynamics of the phase field $\phi$ is defined in terms of the relaxational dynamics
\begin{equation}
	\tau \frac{\partial \phi}{\partial t} = -\frac{\delta E }{\delta \phi},\label{eq:relaxation_dynamics}
\end{equation}
where the parameter $\tau$ determines the rate at which total energy is minimized. After performing the functional derivative we find
\begin{equation}
	\tau \frac{\partial \phi}{\partial t} = K \nabla^2 \phi - f'(\phi).
\end{equation}
This equation describes the dynamics of the phase field $\phi$ and is also known as the Allen-Cahn equation \cite{allen_1979,kobayashi_2010}.

\subsection{Phase field method for morphogen-controlled interface growth}
The dynamics of the phase field corresponding to the sharp interface model presented in the main text is given by
\begin{equation}
	\tau\left( \frac{\partial \phi}{\partial t} + \vecb{u}\cdot\vecb{\nabla}\phi \right) = K  \nabla^2 \phi + r(\phi),\label{eq:app_phase_field_equation_comoving}
\end{equation}
where we have introduced the source term
\begin{equation}
r(\phi) = 2a^2 \phi(1-\phi)\left(\phi-\frac{1}{2}+\frac{\hat{\chi}}{2a^2}\right).
\end{equation}
The parameters $K$, $\tau$, $a$ as well as the bias term $\hat{\chi}$ define the interface properties such as width and velocity. We use a bias term of the form $\hat{\chi} = \hat{\Gamma}\hat{\Theta}$ with
\begin{subequations}
\begin{align}
	\hat{\Gamma}(c) &= \hat{v}_0 - \hat{\gamma} c(x,y)\\
	\hat{\Theta} &= 1 - \delta \left[1 -  \vecb{n} \cdot \vecb{m} \right].
\end{align}
\end{subequations}
and the definition of the interface normal vector
\begin{equation}
    \vecb{n} = -\frac{ {\bm \nabla} \phi }{ |{\bm \nabla} \phi |}.
\end{equation}
For this choice of bias term the phase field dynamics can be shown to recover the interface dynamics of the main text. 

The dynamics of the morphogen concentration $c$ is given by
\begin{equation}
\begin{split}
    \frac{\partial c}{\partial t} +  \vecb{u}\cdot\vecb{\nabla}c &= D \nabla^2 c  - [k(\phi)\\
    &+g_x + g_y]c  + s(\phi).\label{eq:app_morphogen_equation_comoving}  
\end{split}
\end{equation}
Here, we use the phase field dependent degradation and production rate
\begin{subequations}
\begin{align}
	k(\phi) &= \kin \phi  +\kout (1-\phi)\\
	s(\phi) &= \ssin \phi + \ssout (1-\phi).
\end{align}
\end{subequations}
In the next section we rigorously show how the phase field model recovers the sharp interface limit presented in the main text.

To simplify the numerical solution of the phase field model, we transform Eq.~\eqref{eq:app_phase_field_equation_comoving} and Eq.~\eqref{eq:app_morphogen_equation_comoving} from a growing to a non-growing reference frame. To this end, we make the transformation
\begin{equation}
	\begin{split}
		(x,y,t)&\rightarrow (\tildex,\tildey,\tildet)\\
		\tildex &= x e^{-G_x(t)}\\
		\tildey &= y e^{-G_y(t)}\\
		\tildet &= t,
	\end{split}\label{eq:transformation_non_growing}
\end{equation}
where we use tilde to denote quantities in a non-growing domain and introduce the integrated growth rate
\begin{equation}
	G_i(t)= \int_0^t g_i(t') \, d t'.\label{eq:integrated_growth_rate}
\end{equation}

Here, we demonstrate how the phase field and derivatives of it transform while the relations for the morphogen concentration follow analogously. The relation between phase field in the growing and non-growing reference frame is given by
\begin{equation}
	\begin{split}
		\phi(x,y,t) &= \phi(\tildex e^{G_x(t)},\tildey e^{G_y(t)},t)\\
		&= \tildephi(\tildex,\tildey,\tildet).
	\end{split}\label{eq:app_phase_field_transformed}
\end{equation}
From this, we can find the relation between derivatives in growing and non-growing domain by applying the chain rule. We find
\begin{equation}
		\frac{\partial \phi}{\partial t} = -g_x x e^{-G_x(t)}\frac{\partial \tildephi}{\partial \tildex} - g_y y e^{-G_y(t)}\frac{\partial \tildephi}{\partial \tildey} + \frac{\partial \tildephi}{\partial \tildet}.\label{eq:app_time_derivatives_transformed}
\end{equation}
for the time derivative and
\begin{align}
\frac{\partial \phi}{\partial x} &= e^{-G_x(t)}\frac{\partial \tildephi}{\partial \tildex} &
\frac{\partial}{\partial x} \frac{\partial \phi}{\partial x} &= e^{-2 G_x(t)} \frac{\partial^2 \tildephi}{\partial \tildex^2}. \label{eq:app_2nd_derivatives_transformed}
\end{align}
for first and second order spatial derivatives. By combining Eq.~\eqref{eq:app_phase_field_equation_comoving} and Eq.~\eqref{eq:app_morphogen_equation_comoving} together with Eq.~\eqref{eq:app_time_derivatives_transformed} and Eq.~\eqref{eq:app_2nd_derivatives_transformed} we then find that the phase field model in a non-growing reference frame is
\begin{subequations}
\begin{align}
	\tau\frac{\partial \phi}{\partial t}  &= K e^{-2G_x(t)} \frac{\partial^2 \phi}{\partial x^2} + K e^{-2G_y(t)} \frac{\partial^2 \phi}{\partial y^2} + r(\phi)\\
 \begin{split}
     	\frac{\partial c}{\partial t} &= De^{-2G_x(t)} \frac{\partial^2 c}{\partial x^2} + De^{-2G_y(t)} \frac{\partial^2 c}{\partial y^2}
      \\&- [k(\phi)+g_x(t) + g_y(t)]c  + s(\phi),
 \end{split}
\end{align}
\end{subequations}
where we dropped the tilde symbol for simplicity. Note that that irrespective of reference frame the phase field and morphogen dynamics describe the same physical phenomenon, but correspond to a different point of view. In the growing reference frame, internal length scales (e.g., diffusion degradation length) maintain their size, but the system size increases. By contrast, in the non-growing reference frame, internal length scales decrease, while the system size is constant.

\subsection{Orientation field}
In our approach to study branching morphogenesis, we take into account external guiding cues in a coarse grained way in the form of an orientation field $\vecb{m}$. In principle, various environmental influences can act as external guiding cues such as planar cell polarity of mechanical constraints imposed from muscle fibers. Here, we consider a morphogen that is produced on the organism boundary and forms gradient towards the system center as a guiding cue.
We denote the morphogen concentration by $\cext$ and define the corresponding orientation field $\vecb{m}_{\textrm{ext}}$ as
\begin{equation}
	\vecb{m}_{\textrm{ext}} = \frac{\vecb{\nabla}\cext}{|\vecb{\nabla}\cext|}\label{eq:mext}
\end{equation}
The orientation field $\vecb{m}_{\textrm{ext}}$ points in direction of steepest increase in morphogen concentration and allows us to enforce branch growth in this direction.

The dynamics of $\cext$ is governed by the advection-diffusion equation
\begin{align}
\begin{split}
    	\partial_t \cext + \vecb{u} \cdot \vecb{\nabla} \cext = D^x_{\textrm{ext}} \partial^2_x \cext + D^y_{\textrm{ext}} \partial^2_y \cext \\
     - (k_{\textrm{ext}} + g_x + g_y) \cext
\end{split}
 \label{eq:dynamics_external_morphogen1}
\end{align}
together with the boundary condition $\cext|_{\partial \Omega}=\textrm{const}$ which accounts for morphogen production on the organism boundary. We denote the morphogen diffusion constant in the respective direction by where $D^i_{\textrm{ext}}$ and the morphogen degradation rate by $\kext$. The interplay of morphogen diffusion and degradation leads to morphogen gradients with the characteristic diffusion-degradation length $\lambdaext^i=\sqrt{D^i_{\textrm{ext}}/k_{\textrm{ext}}}$ in the respective direction. The diffusion-degradation lengths allow us to consider orientation fields in horizontal ($\lambdaext^x\ll\lambdaext^y$) or vertical direction ($\lambdaext^y\ll\lambdaext^x$).

To solve the equation for external morphogen along the dynamical equation for phase field and morphogen concentration, we also employ transformation from resting to co-moving frame given by Eq.~\eqref{eq:transformation_non_growing} and find that the dynamics of the external morphogen concentration is given by
\begin{align}
\begin{split}
    	\partial_t \cext = D^x_{\textrm{ext}} e^{-2G_x} \partial^2_x \cext + D^y_{\textrm{ext}} e^{-2G_y} \partial^2_y \cext \\
     - (\kext + g_x + g_y) \cext.
\end{split}
 \label{eq:dynamics_external_morphogen2}
\end{align}
This equation describes the formation of a morphogen gradient in a growing organism. For simplicity, we here consider a special case of this dynamics. We consider a scenario in which the diffusion-degradation length scale $\lambdaext^i$ is increased along with organism size as $\lambdaext^i\propto L_i$. Due to the relation $\lambdaext^i\propto e^{G_i}$ the time dependent coefficients in cancel and we are left with an equation with time independent coefficients.
We additionally assume that the dynamics of the external morphogen is faster than any other dynamics and therefore study the case $\partial_t \cext = 0$.
Due to strong horizontal orientation of planarian gut branches, we consider the case of a large gradient in $x$-direction ($\lambdaext^x\ll\lambdaext^y$).

\subsection{Sharp interface limit}
\begin{figure*}
    \centering
    \includegraphics{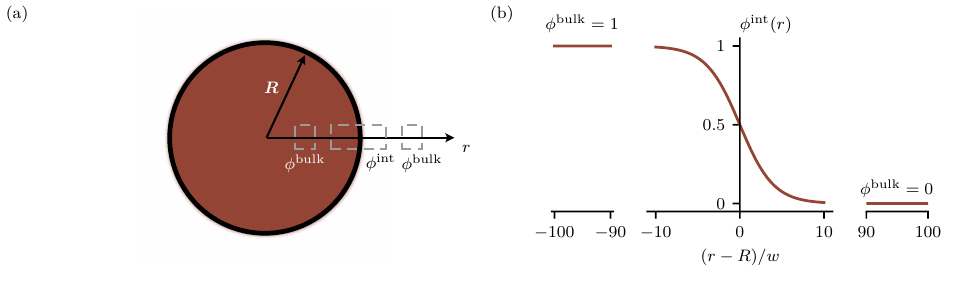}
    \caption{\textbf{Representation of interfaces using a phase field.} (a) We show a phase field (brown) along with its interface (black). We use polar coordinates to describe the structure and denote the radial coordinate by $r$ and the interface position by $R$. (b) We show the phase field $\phibulk$ in the bulk region given by Eq.~\eqref{eq:phase_field_bulk_region} and the phase field $\phiint$ near the interface given by Eq.~\eqref{eq:phase_field_interface_region}.}
    \label{fig:sharp_interface_limit_figure}
\end{figure*}

In this section, we derive the dynamics of the sharp interface model presented in the main text and from dynamics of the phase field model presented in the appendix. For the derivation we adopt polar coordinates with radius $r$ and consider for simplicity a circular structure with radius $R$ described by the phase field $\phi$ (Fig.~\ref{fig:sharp_interface_limit_figure}a). We define the interface width $w=\sqrt{K}/a$. In our derivation we consider the limit of small interface width $w$ ($w/R\ll1$, $w/\lambdai\ll1$) while $\sigma=K/(6w)$ and $\nu=\tau/K$ are constant. Our derivation follows the approach outlined in \cite{elder_2001,provatas_2010,godreche_1991}. 

\subsubsection{Bulk region}
We first obtain the value $\phibulk$ and $\cbulk$ of the phase field and morphogen concentration in the bulk region far away from the interface, respectively. To this end, we first express the Laplacian in Eq.~\eqref{eq:app_phase_field_equation_comoving} by using polar coordinates. We further replace parameters by the interface width $w$ or otherwise constant parameters and find:
\begin{align}
\begin{split}
    \nu w^2\left[\partial_t \phi + \vecb{u}\cdot\vecb{n}\,\partial_r \phi\right] =\\ w^2 \partial^2_{r} \phi + \frac{w^2}{r}\partial_{r}\phi - \frac{1}{2}f_{\textrm{s}}'(\phi) - \frac{w}{6\sigma}\frac{\hat{\chi}}{6}f_{\textrm{t}}'(\phi)\label{eq:app_phase_field_equation_rewritten}
\end{split}
\end{align}
For small interface width $w$ and far away from the interface such that gradients vanish, this equation reduces to
\begin{equation}
	f_{\textrm{s}}'(\phibulk) = 0.\label{eq:phase_field_bulk_region}
\end{equation}
Clearly, this equation has the solutions $\phi=0,1$ giving the phase field in the ``in'' ($\phibulk=1$) and ``out'' region ($\phibulk=0$).
To determine the morphogen field $\cbulk$ in the bulk region, we insert $\phibulk=0,1$ into Eq.~\eqref{eq:app_morphogen_equation_comoving} and find
\begin{align}
    \begin{split}
        \partial_t \cbulk + \vecb{u}\cdot\vecb{\nabla}\cbulk = \\
        D \nabla^2 \cbulk - [k(\phibulk)+g_x +g_y]\cbulk + s(\phibulk).
    \end{split}
\end{align}
This equation reduces to the morphogen dynamics in Eq.~\eqref{eq:morphogen_dynamics} of the main text for the values $\phibulk=0,1$.

\subsubsection{Interface region}
To study the behaviour of phase field and morphogen concentration at the interface, we use the transformation
\begin{equation}
	\tilder = \frac{r - R}{w},\label{eq:transformation_comoving_rescaling}
\end{equation}
where $\tilder$ denotes the radial coordinate in the transformed coordinate frame. This transformation moves the reference frame to the position $R$ of the interface and rescales (``stretches'') positions by the interface width $w$. Applying this transformation to the phase field dynamics in Eq.~\eqref{eq:app_phase_field_equation_comoving} gives
\begin{align}
\begin{split}
    	\nu w^2\left[\partial_t \phi - v_n^{\textrm{tot}} \frac{1}{w}\partial_{\tilder} \phi + \vecb{u}(w\tilde{\vecb{r}}+\vecb{R})\cdot\vecb{n}\frac{1}{w}\partial_{\tilder}\phi\right] = \\
 \partial^2_{\tilder} \phi + \frac{w}{w\tilder+R}\partial_{\tilder}\phi - \frac{1}{2}f_{\textrm{s}}'(\phi) - \frac{w}{6\sigma}\frac{\hat{\chi}}{6}f_{\textrm{t}}'(\phi),
\end{split}
\end{align}
where we have defined $v_n^{\textrm{tot}} = \vecb{n}\cdot \partial_t \vecb{R}$. In the limit $w\rightarrow0$, we find for the phase field $\phiint$ near the interface
\begin{equation}
	0 = \partial^2_{\tilder} \phiint - \frac{1}{2}f_{\textrm{s}}'(\phiint).\label{eq:phase_field_int}
\end{equation}
To ensure that the phase field values match the bulk values, we additionally employ the boundary conditions $\phiint(-\infty)=1$ and $\phiint(+\infty)=0$. The phase field profile near the interface is given by the solution of Eq.~\eqref{eq:phase_field_int} and reads
\begin{equation}
	\phiint(\tilder) = \frac{1}{2}\left[1 - \tanh\left(\frac{{\tilder}}{2}\right) \right].\label{eq:phase_field_interface_region}
\end{equation}
Therefore the phase field value in the ``in'' region and the ``out'' region are connected by a smooth sigmoidal function.

Similarly, we apply the transformation in Eq.~\eqref{eq:transformation_comoving_rescaling} to the morphogen dynamics in Eq.~\eqref{eq:app_morphogen_equation_comoving} and find
\begin{equation}
\begin{split}
	\frac{w^2}{D}\left[\partial_t c - v_n^{\textrm{tot}} \frac{1}{w}\partial_{\tilder} c +  \vecb{u}(w\tilde{\vecb{r}}+\vecb{R})\cdot\vecb{n}\frac{1}{w}\partial_{\tilder}c \right] = \\\partial_{\tilder}^2 c + \frac{w}{\tilder w + R}\partial_{\tilder} c - \frac{w^2}{D} [k(\phi)+g_x+g_y]c \\+ \frac{w^2}{D}s(\phi).
\end{split}
\end{equation}
In the limit of $w\rightarrow0$, we find for the morphogen concentration $\cint$ near the interface
\begin{equation}
	\partial^2_{\tilder} \cint = 0.
\end{equation}
This equation is solved by $\cint(\tilder)=A \tilder + B$, where $A,B$ denote integration constants. Since the morphogen concentration has to be bounded for $\tilder\rightarrow\pm\infty$, we require $A=0$. Therefore, we find that the morphogen concentration is constant.

\subsubsection{Interface velocity}
To determine the interface velocity $v_n$, we transform the coordinate system to a reference frame that is comoving with the interface. Using the transformation $\tilde{r} = r - R$ we obtain
\begin{align}
\begin{split}
    \nu w^2\left[ \partial_t \phi - v_n^{\textrm{tot}} \partial_r \phi + \vecb{u}(\vecb{R}+\vecb{r})\cdot\vecb{n}\partial_r\phi \right] =\\  w^2 \left[\partial^2_r \phi + \frac{1}{r + R}\partial_r \phi \right] - \frac{w}{6\sigma} f'(\phi).\label{eq:phase_field_around_interface}
\end{split}
\end{align}
Next, we consider a small neighborhood $R-\epsilon\ll r \ll R+\epsilon$ around the interface, where $\epsilon$ is a small length scale that satisfies $w\ll \epsilon \ll \lambdai$. We multiply Eq.~\eqref{eq:phase_field_around_interface} on both sides by $\partial_r \phi$ and integrate the resulting equation over the region $R-\epsilon \ll r \ll R+\epsilon$ finding
\begin{align}
	v_n^{\rm tot} &= \frac{1}{\nu}\left[\frac{\hat{\chi}}{6\sigma}-\frac{1}{R}\right] + \vecb{u}(\vecb{R})\cdot \vecb{n}.\label{eq:interface_equation}
\end{align}
To arrive at this equation, we have assumed $\partial_t \phi= 0$ as we are considering stationary interface motion. We have further used $\int (\partial_r \phi)^2 dr=1/(6w)$ which can be derived from an integration of the phase field profile $\phiint$ near the interface and $\int f'(\phi)\partial_r \phi dr = \hat{\chi}/6$ which can be verified by using the chain rule.

The relation for interface velocity now allows us to relate parameters from the phase field model (with hat) and the sharp interface description presented in the main text (without hat). To this end, we recall the definition of $\hat{\chi}=\hat{\Gamma}\hat{\Theta}$ together with
\begin{subequations}
\begin{align}
	\hat{\Gamma}(c) &= \hat{v}_0 - \hat{\gamma} c(x,y)\\
	\hat{\Theta} &= 1 - \delta \left[1 - \frac{ (-{\bm \nabla} \phi) }{ |{\bm \nabla} \phi |} \cdot \vecb{m} \right]
\end{align}
\end{subequations}
By comparing these relations with the respective relations in the sharp-interface limit, i.e., Eqs. \eqref{eq:interface_motion}, \eqref{eq:interface_motion_details}, \eqref{eq:bias_parameter}, \eqref{eq:morphogen_inhibition}, \eqref{eq:angle_dependence},
we can read off the relations $v_0=\frac{\sqrt{K}/a}{\tau}\hat{v}_0$, $\gamma=\frac{\sqrt{K}/a}{\tau}\hat{\gamma}$, $\beta=\frac{K}{a}$. Additionally, we have used the relation
\begin{equation}
\vecb{n} = \frac{ (-{\bm \nabla} \phi) }{ |{\bm \nabla} \phi |}
\end{equation}
derived in \cite{elder_2001}.
Note, that so far we have studied the interface dynamics without noise. We can introduce noise into the description by adding the term $\hat{b}\Lambda$ to the bias parameter $\hat{\chi}$, where $\Lambda$ denotes uniform noise in the range $[-1/2,1/2]$. We find the correspondence $b=\frac{\sqrt{K}/a}{\tau}\hat{b}$. However, we only use $b\neq 0$, when sampling for branch orientations over many realizations.

\subsection{Numerical implementation}
To obtain numerical solutions of the phase field model, we use the finite difference method \cite{press_1992}. We discretize positions $x$ and $y$ by using a grid with spacing $\Delta x$ and $\Delta y$ and use a step size $\Delta t$ for discretization of time $t$. We use an explicit Euler method for the phase field $\phi$ and an implicit Euler method for the morphogen concentration $c$. While the implicit method is stable irrespective of the chosen grid spacing and time step size, the Euler method is stable only for a specific range of grid spacing and step size. The numerical behavior of the finite difference method is captured by the non-dimensional Fourier numbers
\begin{subequations}
\label{eq:Fourier_numbers}
    \begin{align}
	F^n_{\phi,x} &=  \frac{K}{\tau}\frac{\Delta t}{\Delta x^2}e^{-2 G_x(t_n)} \\
    F^n_{\phi,y} &= \frac{K}{\tau}\frac{\Delta t}{\Delta y^2}e^{-2 G_y(t_n)},
    \end{align}
\end{subequations}
where $n$ indicates the current time step. The Euler method is found to be stable if the sum of Fourier numbers is smaller than a threshold.

The constraints posed on Fourier numbers for numerical stability are an important limitation of our numerical method as for a given grid size, the time step has to be chosen sufficiently small to ensure numerical stability. Note that this constraint can be relaxed due to the time dependency of Fourier number for the case of a growing domain. For a growing domain, the Fourier numbers exponentially decrease due to the non-zero growth rate. To circumvent this, we choose the exponentially increasing step size
\begin{equation}
	\Delta t = F_{\phi,x}^0 \frac{\tau}{K} \Delta x^2 e^{2G(t_n)},
\end{equation}
where $G=\min(G_x,G_y)$ and we have assumed a square grid ($\Delta x= \Delta y$). This leads to the time discretization
\begin{equation}
	t_{n+1} = F_{\phi,x}^0 \frac{\tau}{K} \Delta x^2 e^{2G(t_n)} + t_n.
\end{equation}
This adaptive step size scheme reduces the amount of needed time steps drastically while maintaining numerical stability.

\section{Shape instability of a moving interface}\label{sec:stability_analysis}
To understand the formation of unstable interface patterns in our model, we consider a special case of the general model presented in the main text.
For simplicity, we study interface motion in an infinitely long rectangular system with $x\in[0,L_x]$ and $y\in[-\infty,\infty]$. Additionally, we consider a scenario without organism growth ($\vecb{u}=0$) and external guiding cues ($\delta = 0$). This simplified scenario allows us to study basic principles of pattern formation.

The dynamics of the interface vector $\vecb{R}$ for this simplified case is given by
\begin{subequations}
\label{eq:interface_dynamics_no_growth}
	\begin{align}
		\partial_t {\bm R} &= v_n {\bm n}\\
		v_n &= v_0 - \gamma c(\vecb{R}) - \beta \kappa,
	\end{align}
\end{subequations}
where the normal velocity captures the effect of intrinsic organ growth tendency, the inhibition of interface motion by a morphogen as well as the influence of interface curvature as described in the main text. The morphogen concentration is subject to
\begin{equation}
	\partial_t c_i = D \nabla^2 c_i - k_i c_i + s_i\label{eq:morphogen_dynamics_no_growth}
\end{equation}
together with the boundary condition at the interface
\begin{subequations}
\label{eq:bc_interface_no_growth}
\begin{align}
	\cin({\bm R}) &= \cout({\bm R})\\
	{\bm n}\cdot {\bm \nabla} \cin({\bm R}) &= {\bm n}\cdot {\bm \nabla} \cout({\bm R}).
\end{align}
\end{subequations}
Additionally, we use periodic boundary conditions for the left ($x=0$) and right ($x=L_x$) system boundary and the no-flux boundary conditions
\begin{align}
	\lim_{y\rightarrow \pm\infty} \partial_y \ci(x,y) &= 0
\end{align}
at the top and bottom of the rectangular system.

To study the instability of the interface shape, we use the interface position $\yI$. The variable $\yI$ denotes the interface position along the $y$ coordinate and its dynamics can be derived from the interface dynamics given in Eq.~\eqref{eq:interface_dynamics_no_growth}. In a rectangular system the interface vector $\vecb{R}$ can be written as
\begin{equation}
	\vecb{R}(x,t)  =  x \ex + \yI(x,t) \ey.\label{eq:interface_definition}
\end{equation}
We determine the tangent vector $\vecb{\tau}$ of the interface as $\vecb{\tau}=\partial_x \vecb{R}$ and find the interface normal vector $\vecb{n}$ from the condition $\vecb{\tau}\cdot\vecb{n}=0$ to be
\begin{equation}
	\vecb{n} = \frac{-\partial_x \yI \vecb{e}_x + \vecb{e}_y}{\sqrt{1+ (\partial_x \yI)^2}}.
\end{equation}
Here, $\vecb{e}_i$ denotes the unit vector in the respective direction.
From Eq.~\eqref{eq:interface_definition} it follows that interface position dynamics can be derived by $\partial_t \yI=\partial_t \vecb{R}\cdot\vecb{e}_y$ which gives together with the interface dynamics in Eq.~\eqref{eq:interface_dynamics_no_growth}
\begin{align}
	\partial_t \yI &= v_n\frac{1}{\sqrt{1+ (\partial_x \yI)^2}}.
\end{align}
Together with the morphogen dynamics in Eq.~\eqref{eq:morphogen_dynamics_no_growth}, this equation provides the basis of our analysis. Next, we will find the stationary solution of this system of equations and then study its stability.

\subsection{Stationary state}
\subsubsection{Transformation to moving frame}
The stationary solution of this system is a flat interface moving with velocity $v$ accompanied by a concentration profile $c^*_i(y)$. 
To determine this stationary state, we transform the system from a resting frame to a frame co-moving with the interface with velocity $v$. We use the transformation
\begin{equation}
\begin{aligned}
	(x,y,t)&\rightarrow (\tildex,\tildey,\tildet)\\
\tildex &= x\\
\tildey &= y - v t\\
\tildet &= t,
\end{aligned}
\end{equation}
where the tilde labels quantities in the co-moving frame. Note that from now on, we will consider quantities in the co-moving frame only and therefore omit the tilde for simplicity.

The dynamics of interface and morphogen in the co-moving frame reads
\begin{subequations}
\label{eq:dynamics_moving}
\begin{align}
	\partial_t y_{\rm I}&= v_n\frac{1}{\sqrt{1+ (\partial_x \yI)^2}} -v\label{eq:interface_dynamics_comoving}\\
	\partial_t \ci - v \partial_y \ci &= D \nabla^2 \ci - \ki \ci + \ssi\label{eq:concentration_dynamics_comoving}
\end{align}
\end{subequations}
together with the boundary conditions at the interface
\begin{subequations}
\label{eq:bc_moving}
\begin{align}
	\cin(x,0) &= \cout(x,0)\\
	{\bm n}\cdot {\bm \nabla} \cin(x,0) &= {\bm n}\cdot {\bm \nabla} \cout(x,0).
\end{align}
\end{subequations}
Note that an additional advection term appears in Eq.~\eqref{eq:dynamics_moving} which takes into account the relative motion of interface to resting frame.

\subsubsection{Morphogen concentration}
We first determine the stationary morphogen concentration profile $c^*_i(y)$. To this end, we require $\partial_t c^*_i=0$ in Eq.~\eqref{eq:concentration_dynamics_comoving} and solve the resulting stationary diffusion equation
\begin{equation}\label{eq:chap4_stationary_diffusion_eq}
	0 = \nabla^2 c^*_i + \frac{1}{l}\partial_{y} c^*_i - \frac{1}{\lambdai^2}c^*_i + \frac{1}{\lambdai^2}c_i^0,
\end{equation}
where we have introduced the diffusion length $l=D/v$, the reaction-diffusion length $\lambda_i=\sqrt{D/k_i}$ and a typical concentration $c_i^0=s_i/k_i$ for the respective region $i=\textrm{in},\textrm{out}$. We find that the stationary concentration profile obeys
\begin{equation}\label{eq:stationary_concentration_moving_interface}
	c^*_{\textrm{I}}(y) = \mathcal{A}_i \exp{\left[\pm\frac{y}{\ell_i}\right]} + c_i^0
\end{equation}
together with the constants
\begin{equation}
	\mathcal{A}_i = \mp\Delta c \frac{\ell_i}{\lin+\lout} 
\end{equation}
and the effective diffusion-degradation length $\ell_i$ given by
\begin{equation}
	\frac{1}{\li} =  \pm \frac{1}{2l}+ \sqrt{\frac{1}{4l^2} + \frac{1}{\lambdai^2} }.\label{eq:lengthscale_react_diff}
\end{equation}
It can be easily verified that the solution given by Eq.~\eqref{eq:stationary_concentration_moving_interface} satisfies both the stationary diffusion equation as well as all of the boundary conditions.

From the general solution of the morphogen concentration profile we can easily determine the concentration value $c^*_{\textrm{I}}$ at the interface position. To this end, we set $y=0$ in Eq.~\eqref{eq:stationary_concentration_moving_interface} and find
\begin{align}
		c^*_{\textrm{I}}(v) = \Delta c \frac{1}{1+\ratiolilo} + \cout^0\myspace.\label{eq:stationary_concentration_moving_interface_v}
\end{align}
Having determined the concentration value at the interface position, we use this quantity in the next section to determine the interface velocity $v$.

\subsubsection{Interface velocity}
To determine the stationary interface velocity of a flat interface ($\partial_x y^*_{\textrm{I}} =0$) we require $\partial_t y^*_{\textrm{I}}=0$ leading to the equation
\begin{equation}
    c^*_{\textrm{I}}(v) = \frac{v_0}{\gamma} - \frac{1}{\gamma}v.
\end{equation}
This equation constitutes a nonlinear relation that we need to solve to find the stationary interface velocity $v$. Graphically speaking, we are seeking the intersections of the morphogen concentration at the interface with a linear function with slope $- \frac{1}{\gamma}$ and intercept $\frac{v_0}{\gamma}$. The slope of the morphogen concentration determines the number of intersections: 
\begin{subequations}
\begin{align}
	\frac{d c^*_{\textrm{I}}}{dv}\bigg\rvert_{v=0} &<-\frac{1}{\gamma}: \text{up to three solutions}\\
	\frac{d c^*_{\textrm{I}}}{dv}\bigg\rvert_{v=0} &\ge -\frac{1}{\gamma}: \text{exactly one solution}
\end{align}
\end{subequations}
The critical value for when transition takes place can be determined from the condition $\frac{d c^*_{\textrm{I}}}{dv}\bigg\rvert_{v=0}=-1/\gamma$ giving the condition
\begin{equation}
	\Delta c_{\rm crit} = \frac{2D}{\gamma} \left(\frac{1}{\lambdain} + \frac{1}{\lambdaout}\right).
\end{equation}
Finally, we numerically determined the interface velocity $v$ for each of the cases numerically. Note, that we consider only regimes with one solution for $v$.

\subsection{Stability analysis}

\subsubsection{Linearization at the stationary solution}
We next linearize the dynamical equations together with the respective boundary conditions around the stationary state. To this end, we write 
\begin{subequations}
\begin{align}
	\yI(x,t) &= \delta \yI(x,t)\\
	\ci(x,y,t)&= c^*_i(y) + \delta \ci(x,y,t),
\end{align}
\end{subequations}
where $\delta \yI$ denotes a small perturbation around the stationary interface position $y^*_{\textrm{I}}=0$ and $\delta \ci$ denotes a small perturbation around the stationary concentration value $c^*_i$. 
Using this ansatz, we then find for the linearized dynamical equations
\begin{subequations}
\label{eq:dynamics_linearized}
\begin{align}
	\partial_t \delta \yI &= - \gamma [\partial_y c^*(0)\delta \yI + \delta c(0)]  + \beta \partial^2_x \delta \yI \label{eq:dynamics_yI_linearized}\\
	\partial_t \delta \ci &= D {\nabla^2} \delta \ci + v \partial_y \delta \ci - \ki \delta \ci \label{eq:dynamics_ci_linearized}
\end{align}
\end{subequations}
together with the linearized boundary conditions at the interface position
\begin{subequations}
\label{eq:linearized_bc_comoving}
\begin{align}
	\delta \cin(x,0)&=\delta \cout(x,0)\\
\begin{split}
\partial^2_y c^*_{\textrm{in}}(x,0) \delta \yI + \partial_y \delta \cin(x,0) &= \\
\partial^2_y c^*_{\textrm{out}}(x,0) \delta \yI + \partial_y \delta \cout(x,0).
\end{split}
\end{align}
\end{subequations}
The linearized dynamical equations together with the linearized boundary conditions constitute a set of coupled differential equations that we solve next.

\subsubsection{Determination of growth rates}
We solve the set of coupled linearized equations Eq.~\eqref{eq:dynamics_linearized} with the ansatz
\begin{subequations}
\begin{align}
	\delta \yI(x) &= \mathcal{A}_q e^{i q x + \mu t}\label{eq:ansatz_interface_perturbation}\\
	\delta c_i(x,y) &= \delta c_{i,q}(y) e^{i q x + \mu t}.\label{eq:ansatz_concentration_perturbation}
\end{align}
\end{subequations}
Here, $\mathcal{A}_q$ and $\delta c_{i,q}$ denote amplitudes of the interface and morphogen concentration, respectively. We further introduced the growth rate $\mu$ of interface perturbations. According to our convention, $\mu>0$ corresponds to unstable interface growth and $\mu<0$ corresponds to stable interface growth.

To determine the concentration $\delta c_{i,q}$, we insert ansatz for concentration perturbation Eq.~\eqref{eq:ansatz_concentration_perturbation} into the linearized concentration dynamics Eq.~\eqref{eq:dynamics_ci_linearized}. From this approach, we find 
\begin{subequations}
\label{eq:concentration_amplitude}
	\begin{align}
		\delta c_{\textrm{in},k} &= c^{\textrm{in}}_k e^{\qin y}\\
		\delta c_{\textrm{out},k}&= c^{\textrm{out}}_k e^{-\qout y},
	\end{align}
\end{subequations}
where $\qin$ and $\qout$ are the positive solution of the equations
\begin{subequations}
	\begin{align}
		0 &= (\lambdain \tilde{q})^2  - \frac{v}{D}\lambdain^2 \tilde{q} - \left((\lambdain q)^2 + 1 + \frac{\mu}{\kin}\right)\\
		0 &= (\lambdaout \tilde{q})^2  + \frac{v}{D}\lambdaout^2 \tilde{q} - \left((\lambdaout q)^2 + 1 + \frac{\mu}{\kout}\right).
	\end{align}
\end{subequations}
Our sign convention ensures that solutions remain finite for $y=\pm\infty$. We determine the coefficients $c^{\rm in}_q$ and $c^{\rm out}_q$ by inserting the ansatz Eq.~\eqref{eq:ansatz_concentration_perturbation} together with Eq.~\eqref{eq:concentration_amplitude}  into the linearized boundary conditions and find
\begin{subequations}
	\label{eq:amplitude_xi_k}
	\begin{align}
		c^{\rm in}_q &= c^{\rm out}_q \label{eq:chap4_bc_interface_eigenmode} \\
		\mathcal{A}_q [\partial_y^2 c^*_{\textrm{in}} - \partial_y^2 c^*_{\textrm{out}}] &= -\qout c^{\rm out}_q - \qin c^{\rm in}_q.
	\end{align}
\end{subequations}
From this equations, we can determine $c^{\rm in}_q$ and $c^{\rm out}_q$ and thereby fully specify the perturbation $\delta c$.

To finally determine the growth rates $\mu$ we insert Eq.~\eqref{eq:ansatz_interface_perturbation} into the linearized interface dynamics in Eq.~\eqref{eq:dynamics_yI_linearized} and find
\begin{align}
	\mathcal{A}_q \mu = - \gamma [\partial_y c^* \xi_q + \delta c(x,0)] - \beta q^2 \mathcal{A}_q.
\end{align}
We eliminate the interface amplitude $\mathcal{A}_q$ by using Eq.~\eqref{eq:amplitude_xi_k} and further make use of the stationary morphogen concentration profile given in Eq.~\eqref{eq:stationary_concentration_moving_interface} to find for the growth rate
\begin{equation}
	\mu = f(\mu)- \beta q^2, \label{eq:implicit_eq_mu}
\end{equation}
where we have introduced
\begin{widetext}
    \begin{equation}
	f(\mu) = \gamma\Delta c \frac{1}{\lout + \lin}\left[1-\frac{\sqrt{1+ 4 \left(\frac{\ell}{\lambdaout}\right)^2} + \sqrt{1 + 4 \left(\frac{\ell}{\lambdain}\right)^2}}{\sqrt{1 + 4\ell^2\left(\frac{\mu + \kout}{D} + q^2\right)} +\sqrt{1 + 4\ell^2\left(\frac{\mu + \kin}{D} + q^2\right)}}\right].
\end{equation}
\end{widetext}
Eq.~\eqref{eq:implicit_eq_mu} provides an implicit relation for the growth rates $\mu$ that we analyze next.

\subsubsection{Analysis of the growth rates}
To analyze the interface growth rates, we consider the limit of quasistatic morphogen dynamics. In this limit, the morphogen concentration adapts instantaneously to any interface morphology changes. To find the growth rate $\mu$ in this limit, we perform a Taylor expansion of Eq.~\eqref{eq:implicit_eq_mu} for small values of $\mu_k$ and find
\begin{equation}
    \mu =  f(\mu=0) - \beta q^2 + \mu f'(\mu=0) + \mathcal{O}(\mu^2).\label{eq:implicit_eq_mu_linearized}
\end{equation}
Here, we have used
\begin{equation}
f'(\mu=0) = -\frac{\gamma \Delta c}{\lin \lout}\frac{2\ell^2}{D}\frac{1}{(\hin + \hout)^2} \left(\frac{1}{\hin} + \frac{1}{\hout} \right),
\end{equation}
together with the abbreviation
\begin{equation}
    h_i(q) = \sqrt{1 + 4\ell^2\left( \frac{1}{\lambda^2_i} + q^2 \right)}.
\end{equation}
We can then solve Eq.~\eqref{eq:implicit_eq_mu_linearized} for $\mu$ and find
\begin{equation}
    \mu = \frac{f(\mu=0)}{1-f'(\mu=0)} - \frac{\beta q^2}{1-f'(\mu=0)}\label{eq:mu_quasistatic}
\end{equation}
in the quasistatic limit. In contrast to the general relation of interface growth rates Eq.~\eqref{eq:implicit_eq_mu}, this equation provides an explicit relation for the growth rates and allows us to discuss several key features in detail.

In particular, we can determine the critical inhibition parameter value $\gamma_c$ at which the transition from unstable to stable behavior takes place. To this end, we expand Eq.~\eqref{eq:mu_quasistatic} in a Taylor series around $q=0$ up to second order. We find that zeroth and first order term are zero, leaving the second order term as the only non-zero term in our expansion. The sign of the second order term determines whether the parabola is open upwards or downwards and therefore whether interface motion is unstable or stable. We can determine the critical inhibition parameter value from $d^2 \mu/dq^2\rvert_{q=0}=0$ and find
\begin{equation}
	\gamma_c = \frac{\beta}{\Delta c} \frac{\lin + \lout}{2 \ell^2}\left(\frac{2\ell}{\lin}-1\right)\left(\frac{2\ell}{\lout}+1\right).
\end{equation}

By using the quasistatic limit, we can further determine characteristic length scales of this growth process.
For the unstable regime, the simplified equation allows us to derive an approximate relation for the wavevector with maximal growth rate. In the limit $q_{\textrm{max}} \ell\gg1$ (limit of small velocity) and $q_{\textrm{max}}\lambda_i\gg1$ (limit of small branch distance) we find that the wavevector $q^3_{\textrm{max}}$ with maximal growth rate is
\begin{equation}
	q^3_{\textrm{max}} = \frac{\gamma \Delta c}{\beta}\frac{1}{4\lin \lout},\label{eq:lambdamax}
\end{equation}
where we have also used Eq.~\eqref{eq:lengthscale_react_diff}.
Similarly, we find that the wavevector $q_c$ for which the growth rate is zero is
\begin{equation}
	q^2_c = \frac{\gamma \Delta c}{\beta} \frac{1}{\lout + \lin}.
\end{equation}
We can compare the result for the wavevector at which the interface growth rate $\mu$ is zero with the respective relation for other shape instabilities such as the Saffman-Taylor instability \cite{saffman_1958}. In both cases, the characteristic wave length is proportional to the strength of surface tension effects and a difference in material properties is needed for the instability to take place. In the limit of infinitely large reaction diffusion length we additionally recover the inverse proportionality of $\lambda_c$ with interface velocity as for the Saffman-Taylor instability. Note that a similar argument holds for the Mullins-Sekerka instability \cite{langer_1980,mullins_1963}. This discussion explains the relation between the shape instability in morphogen-controlled interface growth and viscous fingering and how similar concepts can govern the formation of branched patterns in different systems.

\begin{figure}
    \centering
    \includegraphics{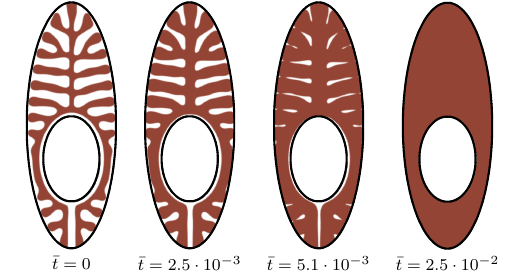}
    \caption{\textbf{Interface dynamics of branched patterns for morphogen inhibition.} We show branched patterns as a function of time $\bar{t}=t/\mu_{\textrm{max}}$ without the influence of the morphogen ($c_i=0$) in a non-growing domain starting from an already established branched pattern.}
    \label{fig:morphogen_inhibition}
\end{figure}
\section{Further parallels between model and the planarian gut}
\subsection{Morphogen inhibition}
To understand the effect of the loss of morphogen in our system, we consider an already established branched pattern in non-growing domain and study its interface dynamics while $c_i=0$ is enforced (Fig.~\ref{fig:morphogen_inhibition}). Due to the loss of morphogen we also observe a loss in mutual branch inhibition. Branches increase in thickness until they fuse and a leaf-like structure forms.

\subsection{Branch formation mechanism in model and experiment}
\begin{figure*}
    \centering
    \includegraphics{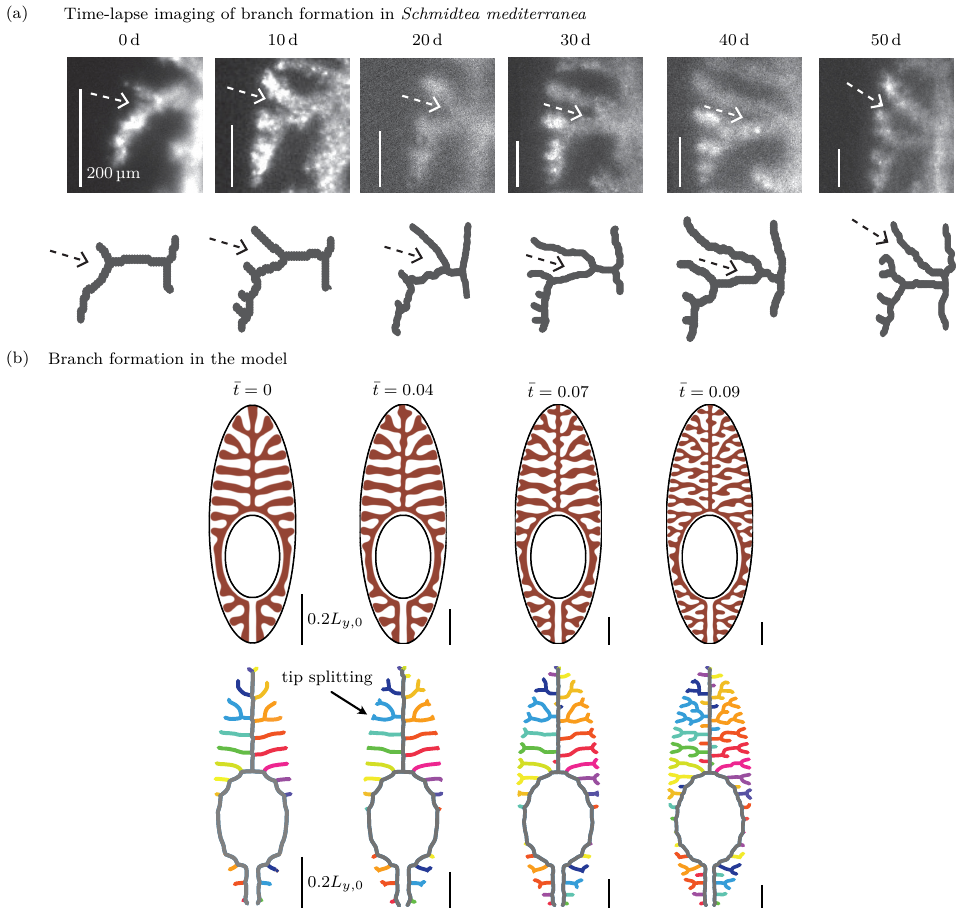}
    \caption{\textbf{Branch formation mechanisms in experiment and model.} (a) We show close ups of a worm fed with fluorescent dextrans 1–2 days prior to imaging at 0, 10, 20, 30, 40, and 50 days. We fed calf liver every 2-3 days between dextran feedings. (b) We show the phase field and the corresponding skeleton at different time points $\bar{t}=t/\mu_{\textrm{max}}$ of branched pattern formation. The color code indicates groups of branches and helps to track branch formation.}
    \label{fig:branch_formation}
\end{figure*}

To understand the mechanisms by which new branches emerge, we study the temporal dynamics of gut formation in \textit{S. mediterranea}. We find that new branches can emerge from tip-splitting and a subsequent unzipping (Fig.~\ref{fig:branch_formation}a). Note that these findings are in agreement with the observations presented in \cite{forsthoefel_2011}.
To study branch formation dynamics in the model, we consider a representative numerical solution (Fig.~\ref{fig:branch_formation}b). As for the experiments, we find that new branches form from tip splitting with subsequent unzipping.

\subsection{Periodic organism (de)growth}\label{app:periodic_degrowth}
\begin{figure}
    \centering
    \includegraphics{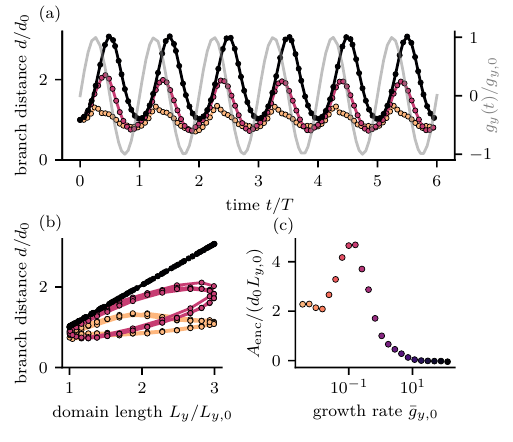}
    \caption{\textbf{Periodic degrowth of simulated gut structures.} (a) Branch distance of simulated gut structures subject to a sinusoidal organism growth rate $g_y(t)$ with period $T$ and amplitude $g_{y,0}$ (gray). $d_0$ denotes the branch distance at time $t=0$. (b) Branch distance as a function of organism length $L_{y,0}$ for different growth rates $g_{y,0}$. (c) Area enclosed by the orbits shown in (b) for different organism growth rates $\bar{g}_{y,0}=g_{y,0}/\mumax$.}
    \label{fig:periodic_degrowth}
\end{figure}

To study the behavior of branched patterns subject to periodic domain growth, we consider a periodic growth rate $g_i(t)$ for $i\in\{x,y\}$. For simplicity, we consider a sinusoidal growth rate $g_i(t) = g_{i,0}\sin(2 \pi t/T)$, where $g_{i,0}$ is the growth amplitude and $T$ denotes the growth period. According to Eq.~\eqref{eq:integrated_growth_rate} the domain size obeys
\begin{equation}
L_{i}(t) = L_{i,0} e^{G_i(t)}
\end{equation}
with the integrated growth rate
\begin{equation}
    G_i(t) = \frac{g_{i,0}T}{2\pi}\left[ 1 - \cos\left(2 \pi \frac{t}{T}\right) \right].
\end{equation}
We choose the growth amplitude such that after half a period the system has increased its length by a factor $s$ ($L_y(t=T/2) = s L_{y,0}$). From this we find the relation
\begin{equation}
    T=\frac{\pi}{g_{y,0}}\log(s)
\end{equation}
for the period $T$.

We track several branch properties and here discuss the behavior of branch distance as an example gut property. We find that branched patterns survive many growth and degrowth cycles characterised by branch distance with a growth rate-dependent amplitude and phase (Fig.~\ref{fig:periodic_degrowth}a). In the limit of fast domain growth, we find that branch distance is in phase with the domain length and increases also by a factor $s$. Clearly, in this limit branched patterns are simply scaled along with the domain. In the limit of slow domain growth, branches have enough time to domain size changes and therefore branch distance changes are much less pronounced.

We next studied branch distance as a function of organism length (Fig.~\ref{fig:periodic_degrowth}b). Periodic (de)growth of branched patterns results in closed paths in the $(d,L_y)$ plane with the enclosed area being a measure for the irreversibility of the (de)growth process. For zero enclosed area, the respective branch distance is independent of organism length and the (de)growth process is therefore reversible. By contrast, for non-zero enclosed area, growth and degrowth follow along different paths indicating the irreversibility of the degrowth process. 

Finally, we used the area enclosed by paths in the $(d,L_y)$ plane to quantify irreversibility of the (de)growth process. We define the enclosed area $A_{\textrm{enc}}$ by
\begin{equation}
    A_{\textrm{enc}} = \frac{1}{n - m} \int_{L_m}^{L_n}d(L)dL.
\end{equation}
We show the result of $A_{\textrm{enc}}$ for our simulations in Fig.~\ref{fig:periodic_degrowth}c. In agreement with previous discussion, we find in general non-zero enclosed area and therefore irreversibility of periodic (de)growth. 

\section{Scaling laws of network length $\Xi$ and segment number $N$ from minimal network model}\label{app:network_exponents}
We also obtain scaling relations for the network length $\Xi=A/\xi_y$, and number of segments $N=A/(\xi_x\xi_y)$ using Eqs.~\eqref{eq:alphacase1},~\eqref{eq:alphacase2}, and ~\eqref{eq:alphacaselog} covering different regimes.
For $g_y>r_y$, we have
\begin{subequations}
\begin{align}
     \eta&=\frac{g_x+r_y}{g_x+g_y}, \quad (g_y>r_y) \quad\\
\zeta &= 
    \frac{r_x+r_y}{g_x+g_y}, \quad (g_y>r_y \; , \; g_x>r_x) \quad \\
    \zeta &= \frac{g_x+r_y}{g_x+g_y}, \quad (g_y>r_y \; , \; g_x<r_x).
\end{align}
\end{subequations}
while in the opposite case
\begin{subequations}
\begin{align}
     \eta&= 1, \quad (g_y<r_y) \quad\\
\zeta &= 
    \frac{g_y+r_x}{g_x+g_y}, \quad (g_y<r_y \; , \; g_x>r_x) \quad \\
    \zeta &= 1, \quad (g_y<r_y \; , \; g_x<r_x).
\end{align}
\end{subequations}

For $g_y=r_y$, we have $\Lambda\sim A/\log(A/A_0)$ and 
\begin{subequations}
\begin{align}
     N& \sim  \frac{A^{\frac{g_y+r_x}{g_x+g_y}}}{\log(A/A_0)}, \quad (g_y=r_y, \; , \; g_x>r_x) \quad\\
N& \sim
    \frac{A}{\log(A/A_0)^2}, \quad (g_y=r_y \; , \; g_x=r_x) \quad \\
    N &\sim \frac{A}{\log(A/A_0)}, \quad (g_y=r_y \; , \; g_x<r_x).
\end{align}
\end{subequations}

\section{Parameter values}
\begin{table*}
	\caption{Parameters used in the main text. The square brackets indicate a range of parameters. In Fig.~3b,d,e we used $b/v_0=1.19\cdot10^{-3}$. Fig.~5 shows parameters identical to Fig.~4, but with $g_x/\kout=0.84\cdot10^{-5}$, $g_y/\kout=1.13\cdot10^{-5}$, and $T\kout=2\cdot10^5$.}
	\centering
	\begin{tabular}{cccccccc} \toprule
		  {Parameter} & {Unit} & {Fig.~2c} & {Fig.~2d} & {Fig.~3a} & {Fig.~3b,d,e} & {Fig.~3c,f,g} & {Fig.~4,6} \\ \midrule
		 $L_x$ & $\lambdaout$  & 0.63 & 20.24 & 15.18 & 15.18 & 15.18 & 15.18 \\
		 $L_y$ & $\lambdaout$  & / & 80.95 & 41.89 & 41.89 & 41.89 & 41.89 \\
		 $T$ & $1/\kout$  & / & $1.25\cdot 10^6$ & $2.5\cdot 10^6$ & $2.5\cdot 10^6$ & $2.5\cdot 10^6$ & $[6.52\cdot10^5,2.25\cdot10^3]$ \\ \midrule
		 $v_0$ & $\lambdaout \kout$  & / & $2.98\cdot10^{-4}$ & $2.39\cdot 10^{-4}$ & $2.39\cdot 10^{-4}$ & $2.39\cdot 10^{-4}$ & $2.39\cdot10^{-4}$ \\
		 $\gamma$  & $\lambdaout \kout$  & $[31.6,0,-6.32]\cdot 10^{5}$  & $1.56\cdot10^{-4}$ & $1.49\cdot 10^{-4}$ & $1.49\cdot 10^{-4}$  & $1.49\cdot 10^{-4}$ & $1.49\cdot 10^{-4}$  \\
		 $\beta$  & $\lambdaout^2 \kout$ & $2\cdot10^5$  & $2.12\cdot10^{-5}$ & $1.47\cdot10^{-5}$  & $1.47\cdot10^{-5}$ & $[8.8,35.2]\cdot10^{-6}$ & $1.47\cdot10^{-5}$   \\ 
		 $v$  & $\lambdaout \kout$ & $1.26\cdot10^6$  & / & /  & / & / & /  \\ 
		 $\delta$  & / & /  & 0,0.4 & 0.4  & $[0,0.5]$ & 0.4 & 0.4  \\ \midrule
		 $D$   & $\lambdaout^2 \kout$ & 1  & 1 & 1 & 1 & 1 & 1  \\
		 $\kin$   & $\kout$ & 1   & 1 & 1 & 1  & 1 & 1 \\
		 $\kout$  & $\kout$ & 1  & 1 & 1 & 1  & 1 & 1  \\
		 $\ssin$  & $\kout$ & 3  & 3 & 3.3 & 3 & 3.3 & 3.3  \\
		 $\ssout$  & $\kout$ & 0  & 0 & 0  & 0.1 & 0 & 0  \\ 
		 $g_x$  & $\kout$ & 0  & 0 & 0  & 0 & 0 & $0.75\cdot[3.46\cdot10^{-6},10^{-2}]$ \\  
		 $g_y$  & $\kout$ & 0  & 0 & 0  & 0 & 0 & $[3.46\cdot10^{-6},10^{-2}]$  \\ \midrule
   $D^x_{\textrm{ext}}$ & $\lambdaout^2 \kout$ & / & / & $3.28\cdot10^{-4}$ & $3.28\cdot10^{-4}$ & $3.28\cdot10^{-4}$ & $3.28\cdot10^{-4}$ \\ 
   $D^y_{\textrm{ext}}$ & $\lambdaout^2 \kout$ & / & / & 0 & 0 & 0 & 0\\ 
   $k_{\textrm{ext}}$ & $\lambdaout^2 \kout$ & / & / & $10^{-3}$ & $10^{-3}$ & $10^{-3}$ & $10^{-3}$\\ 
   $\bpb$ & $\lambdaout$ & / & / & 1 & 1 & 1 & 0.5 \\ \midrule
		 $\sqrt{K}/a$   & $\lambdaout$ & / & $3.58\cdot10^{-2}$ & $3.58\cdot10^{-2}$ & $3.58\cdot10^{-2}$ & $3.58\cdot10^{-2}$ & $3.58\cdot10^{-2}$  \\
		 $\frac{\sqrt{K}/a}{\tau}$   & $\lambdaout \kout$ & / & $1.19\cdot10^{-3}$ & $1.19\cdot10^{-3}$ & $1.19\cdot10^{-3}$  & $1.19\cdot10^{-3}$ & $1.19\cdot10^{-3}$ \\ \bottomrule
	\end{tabular}
\vspace{1em}
\caption{Parameters used in the appendix.}
 \begin{tabular}{ccccc} \toprule
		  {Parameter} & {Unit} &  {Fig.~11} & {Fig.~12} & {Fig.~13} \\ \midrule
		 $L_x$ & $\lambdaout$  &  15.18 & 15.18 & 15.18  \\
		 $L_y$ & $\lambdaout$  &  41.89 & 41.89 & 41.89 \\
		 $T$ & $1/\kout$  &  $2.5\cdot 10^6$ & $2\cdot 10^6$ & $[7.38\cdot10^7,2.07\cdot10^3]$ \\ \midrule
		 $v_0$ & $\lambdaout \kout$  &  $2.39\cdot 10^{-4}$ & $2.39\cdot 10^{-4}$  & $2.39\cdot 10^{-4}$  \\
		 $\gamma$  & $\lambdaout \kout$  &  $1.49\cdot 10^{-4}$ & $1.49\cdot 10^{-4}$ & $1.49\cdot 10^{-4}$  \\
		 $\beta$  & $\lambdaout^2 \kout$ &  $1.47\cdot10^{-5}$ & $1.47\cdot10^{-5}$ & $1.47\cdot10^{-5}$  \\ 
		 $v$  & $\lambdaout \kout$ &  0.4 & /  & / \\ 
		 $\delta$  & /  & 0.4 & 0.4  & 0.4 \\ \midrule
		 $D$   & $\lambdaout^2 \kout$  & 1 & 1 & 1 \\
		 $\kin$   & $\kout$ &  1 & 1 & 1  \\
		 $\kout$  & $\kout$ &  1 & 1 & 1  \\
		 $\ssin$  & $\kout$ & 0 & 3.3 & 3.3   \\
		 $\ssout$  & $\kout$   & 0 & 0  & 0 \\ 
		 $g_x$  & $\kout$ &  0 & $0.84\cdot10^{-6}$ & $0.75\cdot[2.81\cdot10^{-7},10^{-2}]$ \\  
		 $g_y$  & $\kout$ &  0 & $1.13\cdot10^{-5}$ & $[2.81\cdot10^{-7},10^{-2}]$ \\ \midrule
   $D^x_{\textrm{ext}}$ & $\lambdaout^2 \kout$ & $3.28\cdot10^{-4}$ & $3.28\cdot10^{-4}$ & $3.28\cdot10^{-4}$  \\ 
   $D^y_{\textrm{ext}}$ & $\lambdaout^2 \kout$ & 0 & 0 & 0 \\ 
   $k_{\textrm{ext}}$ & $\lambdaout^2 \kout$ & $10^{-3}$ & $10^{-3}$ & $10^{-3}$ \\ 
   $\bpb$ & $\lambdaout$ & 1 & 0.5 & 0.5 \\ \midrule
		 $\sqrt{K}/a$   & $\lambdaout$  & $3.58\cdot 10^{-2}$ & $3.58\cdot 10^{-2}$ & $3.58\cdot 10^{-2}$ \\
		 $\frac{\sqrt{K}/a}{\tau}$   & $\lambdaout \kout$  & $1.19\cdot10^{-3}$ & $1.19\cdot10^{-3}$ & $1.19\cdot10^{-3}$ \\ \bottomrule
	\end{tabular}
\end{table*}

\end{document}